\documentclass[aps,a4paper,12pt]{revtex4}
\pdfoutput=1
\usepackage{float}
\usepackage{fancyhdr}
\usepackage{color}
\usepackage{graphicx}
\usepackage{epsfig} 
\usepackage{amssymb}
\usepackage{amsmath,amsfonts}
\usepackage{mathdots}
\usepackage{mathtools}
\usepackage{booktabs}
\usepackage{hyperref}
\usepackage{tikz}
\usetikzlibrary{shapes}
\usepackage{pgfplotstable}
\usepackage[footnotesize,singlelinecheck=off,justification=raggedright]{caption}
\usepackage{multirow}
\def\ket#1{\mathinner{|{#1}\rangle}}

\def\prod#1#2{\mathinner{\langle{\,#1\,}|{\,#2\,}\rangle}}

\DeclareMathAlphabet{\mathbbmsl}{U}{bbm}{m}{sl}
{\catcode`\|=\active\gdef\Braket#1{\left<\mathcode`\|"8000\let|\bravert {#1}\right>}}
\def\bravert{\egroup\,\vrule\,\bgroup}

\def\ii{{\rm i}}
\def\dd{{\rm d}}
\def\ee{{\rm e}}

\def\CH{{\mathcal H}}
\def\CB{{\mathcal B}}
\def\CL{{\mathcal L}}
\def\CK{{\mathcal K}}
\def\CV{{\mathcal V}}
\def\CS{{\mathcal S}}

\def\NC{{\mathbb C}}
\def\NR{{\mathbb R}}
\def\NZ{{\mathbb Z}}
\def\con{q}

\def\pref#1{(\ref{#1})}
\def\trans{{\!\mathfrak{t}}}
\def\comment#1{}
\def\bulk{{\rm b}}
\def\edge{{\rm e}}
\def\vedge{{\rm def}}
\def\lspan{{\rm span}}
\def\peq{{{\scriptscriptstyle <}}}
\def\gra{{{\scriptscriptstyle >}}}
\def\Ker{{\,\rm Ker\,}}
\def\id{{\rm id}}
\newcommand{\Cross}{\mathbin{\tikz [x=1.4ex,y=1.4ex,line width=.1ex] \draw (0,0) -- (1,1) (0,1) -- (1,0) (0, 0.5) -- (1, 0.5); }}%

\begin{document}

\title{Spectrum of localized states in fermionic chains with defect and adiabatic charge pumping.}

 \author{Filiberto Ares}
\email{faresase@sissa.it}
\affiliation{International School for Advanced Studies (SISSA), 34136 Trieste, Italy} 
\affiliation{International Institute of Physics, UFRN, 59078-970 Natal, RN, Brazil}
\author{Jos\'e G. Esteve}
 \email{esteve@unizar.es}
 \affiliation{Departamento de F\'{\i}sica Te\'orica, Universidad de Zaragoza,
50009 Zaragoza, Spain}
\affiliation{Instituto de Biocomputaci\'on y F\'{\i}sica de Sistemas
Complejos (BIFI), 50009 Zaragoza, Spain}   
\affiliation{Centro de Astropart\'\i culas y F\'\i sica de Altas Energ\'\i as (CAPA)
50009 Zaragoza, Spain}
  \author{Fernando Falceto}
\email{falceto@unizar.es}
 \affiliation{Departamento de F\'{\i}sica Te\'orica, Universidad de Zaragoza,
50009 Zaragoza, Spain}
\affiliation{Instituto de Biocomputaci\'on y F\'{\i}sica de Sistemas
Complejos (BIFI), 50009 Zaragoza, Spain}   
\affiliation{Centro de Astropart\'\i culas y F\'\i sica de Altas Energ\'\i as (CAPA)
50009 Zaragoza, Spain}
    


\begin{abstract} 
In this paper, we study the localized states of a generic quadratic fermionic chain
with finite-range couplings and an inhomogeneity in the hopping (defect)
that breaks translational invariance. When the hopping of the defect vanishes,
which represents an open chain, we obtain a simple bulk-edge correspondence: the zero-energy 
modes localized at the ends of the chain are related to the roots of a polynomial determined
by the couplings of the Hamiltonian of the bulk. From this result, we define an index that characterizes the different topological phases
of the system and can be easily computed by counting the roots of the polynomial. 
As the defect is turned on and varied adiabatically, the zero-energy modes may cross 
the energy gap and connect the valence and conduction bands. We analyze the robustness
of the connection between bands against perturbations of the Hamiltonian.
The pumping of states from one band to the other allows the creation of particle-hole pairs
in the bulk.
An important ingredient for our analysis is the transformation of the Hamiltonian under the standard discrete symmetries, $C$, $P$, $T$,
as well as a fourth one, peculiar to our system,  that is related to the existence of a gap and localized states.

  \end{abstract}

\maketitle

\section{Introduction}\label{sec_intro}

Topology has played an outstanding role in condensed matter physics since
the discovery of quantum Hall effect and topological phase transitions 
\cite{Asorey}. Among the systems whose topological 
properties have been widely studied, we find topological insulators 
\cite{Qi, Hasan, Bernevig}.
These are systems of non-interacting 
fermions that possess gapped phases with zero-energy edge states. 
These modes are spatially localized at the boundaries of the system 
and they are topologically protected in the sense that they are robust 
against deformations of the Hamiltonian, as long as the gap of the bulk 
does not close and certain symmetries are preserved. A prominent example 
is the open Kitaev chain, which hosts Majorana modes at its end-points \cite{KitaevMajorana}
that according to some studies may be detected experimentally, see e.g. 
\cite{Rokhinson, Das, Mourik, Nadj, SarmaMaj}.
The robustness of the edge modes under small perturbations or disorder 
has turned topological insulators into essential 
ingredients of quantum devices \cite{KitaevAnyons, Pachos, Sarma}. 

Topological insulators can be classified in terms of the behaviour 
of its Hamiltonian under charge conjugation ($C$), parity ($P$) and time 
reversal ($T$) symmetries. According to the ten-fold classification \cite{Altland, Kitaev, Ryu}, 
these discrete symmetries allow to arrange them in a ``periodic table'' of 
ten generic classes, provided the Hamiltonian is Hermitian, each one 
related to one of the ten Cartan symmetric spaces. A consequence of this 
classification is that topologically inequivalent systems can not be adiabatically
connected without breaking some of the classifying symmetries.
The different phases within a topological insulator can be characterized 
by a topological invariant. The topological invariant typically establishes 
a relation between a property of the bulk and another one of the edge---the 
so-called bulk-edge correspondence \cite{RyuBB}. 

In this paper, we present a comprehensive and systematic analysis of the edge states in 
one-dimensional chains of spinless fermions described by a generic homogeneous 
quadratic Hamiltonian with finite-range couplings. The end-points of the chain are 
connected by a hopping term parametrized by $q$, which will be referred as the contact. 
We may see this special bond as a defect in the chain which breaks the translational
invariance \cite{Henkel, Grimm, Eisler, Bertini, Alase, Cobanera, Reyes-Lega, Najafi, Ares1}. Its existence 
will be very important for us, as it produces, under certain circumstances, localized 
states in the vicinity of the defect. 

In particular, we will analyze the zero modes of the open chain, $q=0$, through the 
discrete symmetries $P$, $C$, and $T$. The dimension of the space of zero-energy 
modes characterizes the universality class of the corresponding Hamiltonian. 
We will see that these modes are associated to the roots of a polynomial defined 
on the analytical continuation of the momentum space to the Riemann sphere and with
coefficients given by the bulk Hamiltonian. This result leads to a simple bulk-edge 
correspondence for these systems and, as we will show, the determination of the 
topological phase reduces to 
a counting of the roots of the aforementioned polynomial. 

We will also study the spectrum of localized states when the defect is turned on, 
$q\neq 0$. We obtain that, when $P$ and $C$ are 
symmetries of the chain, there are localized states that traverse the bulk gap as 
the contact $q$ varies, allowing the pumping of states between the valence and conduction 
bands. In other systems, this mechanism may induce a quantized transport of charge across
the bulk without applying any bias voltage---a genuine topological phenomenon known as adiabatic charge pumping 
\cite{Thouless, Asboth, Kraus, Nakajima, Lohse, Zilberberg, Kuno}---or may switch the fermionic parity of the edge states 
---a fermion parity pump \cite{Teo, Keselman}. In our case, the pumping of a localized state to the 
conduction band allows to create a free particle-hole pair delocalized in the bulk. An important 
difference is that, while in the usual topological pumping the parameter that is 
changed refers to a feature of the bulk, e.g. modulating their hopping amplitudes, here we vary adiabatically the value of the contact that characterizes the defect.

 The rest of the paper is organized as follows. In Sec. \ref{sec_long_range}, we 
 introduce the family of quadratic fermionic systems under study, as well as a 
 formalization of the well-known diagonalization procedure for this kind of systems
 in terms of Bogoliubov modes.
 In Sec.~\ref{sec:symm}, we discuss the discrete symmetries $C$, $P$ and $T$, as 
 they are crucial in the analysis of the edge modes. In fact, they will be used in Sec.~\ref{sec:zero_modes} to determine 
 the number of zero-energy modes present in the open chain. In Sec.~\ref{sec:bogoliubov_states}, we
 describe how to construct the edge states; providing a rigorous definition 
 for them in the thermodynamic limit, and we analyze their basic properties. 
 Using this framework, in Sec.~\ref{sec:bulk_edge}, we rederive in an analytic way 
 the algebraic results of Sec.~\ref{sec:zero_modes} and, moreover, we formulate a bulk-edge
 correspondence and introduce a topological index, which identifies the 
 different phases of this class of systems. Sec.~\ref{sec:pump} is devoted to the analysis of the 
 spectrum of localized states in terms of the contact $q$ and to the charge pumping 
 phenomenon for different topological phases. Finally, in Sec.~\ref{sec:conclusions}, we end with the conclusions
 and future prospects.

\section{Long range fermionic chain. Space of Bogoliubov modes.}
\label{sec_long_range}

We consider a fermionic chain of length $N$.
Its Hilbert space $\CH=(\NC^2)^{\otimes N}$ is endowed
with the standard scalar product.
In this space, the creation and annihilation operators, $a_n^\dagger$ and  $a_n$,
act with anticommutation relations
$$\{a_n,a_m\}=\{a_n^\dagger,a_m^\dagger\}=0,\quad\{a_n,a_m^\dagger\}=\delta_{nm}\id,$$
and $\id$ is the identity operator in $\CH$.

Our system is a homogeneous chain with range $L$ couplings
and quadratic Hamiltonian.
The two ends of the chain are connected with a tight binding type
of interaction with real hopping parameter $\con$, we will call it contact.
This is our main tunable parameter and, in particular, we recover
the open chain by taking $\con=0$.

According to the previous description, the most general form of the
Hamiltonian is
\begin{eqnarray*}
  H&=&\frac12\sum_{n=1}^N\sum_{l=-L}^L\hskip -1mm{\vphantom{\sum}}'
  \left(2 A_l\, a_n^\dagger a_{n+l}+
  B_l\, a_n^\dagger a_{n+l}^\dagger-B'_l\, a_na_{n+l}\right)\\
  &+&\con\,(a_1^\dagger a_N+a_N^\dagger a_1),
\end{eqnarray*}
where $\sum'$ stands for the sum restricted to terms such that $1\leq n+l\leq N$
and, without loss of generality, we may take $B_{-l}=-B_l$,
$B'_{-l}=-B'_l$. We will always assume that $N>2L$.

Of course, in general $H$ is not self-adjoint. In due time, for actual
computations, we will restrict to the self-adjoint case that requires
$$A_{-l}=\overline A_l,\qquad
B_l'=\overline B_l.
$$
However, and in order to study the symmetries of the system, we prefer
for the moment to keep the most general form for $H$. 

The usual strategy in order to find the spectrum of this kind of quadratic
systems is to write the Hamiltonian in terms of the Bogoliubov modes
which render it diagonal~\cite{Lieb}. A convenient way to formalize this procedure
is by the introduction of another Hilbert space, different from $\CH$,
that for the lack of a better name we will call {\it space of Bogoliubov modes} 
and will be denoted by $\CB$. It is simply the space generated by the linear 
span of the creation and annihilation operators, that is
$$\CB=\operatorname{span}\{a^\dagger_1,\dots,a^\dagger_N,a_1,\dots,a_N\}.$$
Note that $\CB$ is a subspace of $\CL(\CH)$, the linear operators in $\CH$.

Using the anticommutation of operators we may introduce a scalar product in
$\CB$, namely 
\begin{equation}\label{scalar_product}
  \prod\xi{\xi'}\id=\{\xi^\dagger,\xi'\},\quad \xi,\xi'\in\CB.
\end{equation}
  One immediately sees that
the standard basis
$\{a^\dagger_1,\dots,a^\dagger_N,a_1,\dots,a_N\}$
is orthonormal with respect to $\prod{\cdot}{\cdot}$.

Now, the Hamiltonian acts naturally in $\CB$ by the adjoint. We denote
by $H_{\CB}$ the corresponding operator, so that
$$H_{\CB}\xi=[H,\xi].$$
Here the right hand side should be understood as the commutator of
operators in $\CL(\CH)$ that, for $\xi\in\CB$ and due to the quadratic nature of
$H$, also belongs to $\CB$. 

The matrix for $H_\CB$ in the standard basis can be written as
$$(H_\CB)=
\begin{pmatrix}\mathbf{A}&\mathbf{B}\\
-\mathbf{B'}&-\mathbf{A}^{\trans}\end{pmatrix},
$$
where $^{\trans}$ stands for transposition and
$\mathbf{A}$, $\mathbf{B}$ are Toeplitz matrices
given by
\small{$$\mathbf{A}=
\begin{pmatrix}
  A_0&A_1&\cdots&A_L&0&\cdots&0&\con\\[-1.5mm]
  A_{-1}&A_0&\ddots&&\ddots&\ddots&&0\\[-1.5mm]
  \vdots&\ddots&\ddots&\ddots&&\ddots&\ddots&\vdots\\[-1.5mm]
  A_{-L}&&\ddots&\ddots&\ddots&&\ddots&0\\[-1.5mm]
  0&\ddots&&\ddots&\ddots&\ddots&&A_L\\[-1.5mm]
  \vdots&\ddots&\ddots&&\ddots&\ddots&\ddots&\vdots\\[-1.5mm]
  0&&\ddots&\ddots&&\ddots&A_0&A_1\\[-1.5mm]
  \con&0&\cdots&0&A_{-L}&\cdots&A_{-1}&A_0
\end{pmatrix},
\quad
\mathbf{B}=
\begin{pmatrix}
  0&B_1&\cdots&B_L&0&\cdots&0&0\\[-1.5mm]
  B_{-1}&0&\ddots&&\ddots&\ddots&&0\\[-1.5mm]
  \vdots&\ddots&\ddots&\ddots&&\ddots&\ddots&\vdots\\[-1.5mm]
  B_{-L}&&\ddots&\ddots&\ddots&&\ddots&0\\[-1.5mm]
  0&\ddots&&\ddots&\ddots&\ddots&&B_L\\[-1.5mm]
  \vdots&\ddots&\ddots&&\ddots&\ddots&\ddots&\vdots\\[-1.5mm]
  0&&\ddots&\ddots&&\ddots&0&B_1\\[-1.5mm]
  0&0&\cdots&0&B_{-L}&\cdots&B_{-1}&0
\end{pmatrix},
$$
}
and $\mathbf{B}'$ is like $\mathbf{B}$ with $B'_l$ replacing $B_l$.

We will use the symbol $^+$ for the adjoint in the Hilbert space
$\CB$ and keep the more standard $^\dagger$ for the adjoint in $\CH$.
Then notice that
$$
(H_\CB)^+=
\begin{pmatrix}\mathbf{A}^+&\overline{\mathbf{B}'}\\
  -\overline{\mathbf{B}}&-(\mathbf{A}^+)^{\trans}
\end{pmatrix},
$$
where we have used the antisymmetry of $\mathbf{B}$ and
$\mathbf{B}'$.
Therefore, $H_\CB$ is Hermitian if and only if $\mathbf{A}^+=\mathbf{A}$
and $\overline{\mathbf{B}'}={\mathbf{B}}$ or, in other words, if and only if
$A_{-l}=\overline A_l$ and $B_l'=\overline B_l$, which are
precisely the conditions for the Hermiticity of $H$.
This can also be proven using the definition of the scalar product
in terms of the anticommutator,
$$\prod\xi{H_\CB\,\xi'}=\{\xi^\dagger,[H,\xi']\}=
\{[\xi^\dagger,H],\xi'\}=
\{[H^\dagger,\xi]^\dagger,\xi'\}.$$
This implies that
$$H_\CB \xi=[H^\dagger,\xi]$$
and we recover the previous result: the Hermiticity of $H$ in $\CH$
is equivalent to that of $H_\CB$ in $\CB$.

Now the Bogoliubov modes are the eigenvectors of $H_\CB$
$$H_\CB\xi_n=-E_n\xi_n$$
and provided $H$ is Hermitian they form an orthogonal basis.

In this case (Hermitian Hamiltonian) there is also the
(anti)symmetry under the adjoint operation that inverts
the sign of the energy for a Bogoliubov mode
$$H_\CB \xi_n^\dagger=[H,\xi_n^\dagger]=-[H,\xi_n]^\dagger=E_n\xi_n^\dagger.$$ 

From all the previous considerations we may divide the Bogoliubov modes
into those with negative energy $\xi_n$ and their adjoint
$\xi^\dagger_n$ with positive energy (for the present consideration we can
forget about the possible states with zero energy, although they will be studied at length later on).
As stated before, they are orthonormal, which, given the definition of the scalar product in (\ref{scalar_product}),
implies that they satisfy canonical anticommutation rules,
$$\{\xi_n^\dagger,\xi_m\}=\delta_{nm}\id,$$
while
$$\{\xi_n,\xi_m\}=0\quad{\rm and}\quad
\{\xi_n^\dagger,\xi_m^\dagger\}=0.$$

Of course, the importance of the Bogoliubov modes is that they
{\it diagonalize} the Hamiltonian in the sense that the latter
can be written in the form
$$H=\sum_{n}E_n\xi_n^\dagger\xi_n,$$
and from this expression one easily finds the
energy of the multiparticle states in the associated
Fock space.

\section{Discrete symmetries}\label{sec:symm}

After this brief formalization of the Bogoliubov procedure for
solving quadratic Hamiltonians, we want to study the discrete
symmetries of the system. Later we will see that they play an important
role in the discussion of the localized states.

{\bf Parity}

The first discrete symmetry that we consider is parity
or space reflection. We represent it with a unitary operator
$P\in\CL(\CB)$ and, if it satisfies $P^2=I$, then it must be also Hermitian.
It implements a spatial reflection with respect to the defect
and therefore it should have the general form
$$
P: a_n\mapsto\eta a_{N+1-n},\quad
P: a^\dagger_n\mapsto\rho a^\dagger_{N+1-n},
$$
where in order to satisfy the requirements of unitarity and Hermiticity
$\rho$ and $\eta$ should be $1$ or $-1$. A global factor of $-1$
is irrelevant, therefore we can always take $\rho=1$. The matrix
for $P$ in the standard basis is
$$
(P)=\begin{pmatrix}
\mathbf J&0\\
0&\eta\mathbf J
\end{pmatrix},
\quad {\rm with}\quad
\mathbf{J}=\begin{pmatrix}
0&\cdots&0&1\\[-3mm]
\vdots&\iddots&\iddots&0\\[-2.5mm]
0&\iddots&\iddots&\vdots\\[-3mm]
1&0&\cdots&0
\end{pmatrix}
$$
the antidiagonal $N\times N$ matrix.
It is immediate to check that $\mathbf{JAJ}=\mathbf{A}^{\trans}$,
$\mathbf{JBJ}=-\mathbf{B}$
and $\mathbf{JB'J}=-\mathbf{B}'$.
Therefore, we have
$$
(PH_\CB P)=\begin{pmatrix}\mathbf{A}^{\trans}&-\eta\mathbf{B}
\\
-\eta\mathbf{B}'&-\mathbf{A}
\end{pmatrix},
$$
and, if we fix the value of $\eta$ to $-1$, then parity is a symmetry of the
Hamiltonian, $PH=HP$,  if and only if $\mathbf{A}^{\trans}=\mathbf{A}$
or $A_l=A_{-l}$.
Notice that, if we restrict to Hermitian Hamiltonians
(that is the case of interest), this is equivalent
to having $A_l\in\NR$.

{\bf Time reversal}

As it is well known time reversal, denoted by $T$, is represented by
  an antiunitary operator and it should be also involutive: $T^2=I$.
  Under these conditions, as proved already by Wigner \cite{Uhlman},
  there is always a basis on which $T$ acts as the
  identity. Therefore, the only effect of the antilinear operator in a
  general vector is to take the complex conjugation of its components
  in that basis.

  We shall assume that the standard basis is left unchanged by $T$.
  Hence $T$ is simply the complex conjugation of the
  coefficients which may be expressed as $(T)=\overline{\phantom A}$,
  where we use the overline for complex conjugation.
  Therefore, $T$ is a symmetry of the Hamiltonian in the sense that
  $TH=HT$ if and only if the couplings are real, $A_l,B_l,B'_l\in\NR$.

  Notice also that, due to the fact that the matrix of $P$ in the standard
  basis has real entries, it commutes with time reversal, $TP=PT$.
  
{\bf Charge conjugation}

The last of the three discrete transformations exchanges
creation and annihilation operators. It will be implemented by a unitary,
involutive linear map that we denote by $C$.

If we call by $b_k$ and $b^\dagger_k$ respectively
the annihilation and creation operators of a particle
with momentum $k$,
$$b_k=\sum_{n=1}^N \ee^{-\ii k n}a_n,\quad
b^\dagger_k=\sum_{n=1}^N \ee^{\ii k n}a^\dagger_n,
$$
we require that the charge conjugation operator transforms
one into the other (up to a phase)
$$C:b_k\mapsto \varphi b_k^\dagger.$$
It is clear that in order to achieve this goal we must combine the
transformation of creation and annhilation operators in the space
representation with a reflection, namely
$$
C:a_n\mapsto \eta a^\dagger_{N-n+1},\quad
C:a_n^\dagger\mapsto \rho a_{N-n+1},
$$
with phases $\eta$ and $\rho$.
This transformation implies that $b_k$ is mapped into
$\ee^{-\ii k(N+1)}\eta b_k^\dagger$
as required.

Now we want $C^+=C^{-1}=C$, therefore the most general form for
$C$ in the standard basis is
$$(C)=\begin{pmatrix}
0&\eta\mathbf J\\
\overline\eta\mathbf J&0
\end{pmatrix}.
$$

The transformed Hamiltonian is given by
$$(CH_\CB C)=-\begin{pmatrix}\mathbf{A}&-\eta^2\mathbf{B}'
\\
{\overline\eta}^2\mathbf{B}&-\mathbf{A}^\trans
\end{pmatrix},
$$
and therefore $C$ reverses the sign of the energy,
\begin{equation}\label{Csymmetry}
  CH_\CB=-H_\CB C,
  \quad\mbox{if and only if}\quad 
  \eta^2 \mathbf{B}'=-\mathbf{B}.
\end{equation} 

So far the value of $\eta$ is undetermined.
It is related to the relative phase between creation and annihilation
operators and can be modified by redefining them. The only effect
of the latter redefinition in the Hamiltonian is to multiply the
coefficients $B_l$ and $B_l'$ by the corresponding phase.
However, and in view of the condition in (\ref{Csymmetry}), it is very
natural to fix $\eta=\ii$ and then we have that
$C$ anticommutes with $H_\CB$ if $\mathbf{B}=\mathbf{B}'$.

If we include the condition for Hermiticity of the Hamiltonian,
$\overline{\mathbf{B}}=\mathbf{B}'$, the result is that an Hermitian
Hamiltonian anticommutes with $C$ if $B_l\in\NR$.

To summarize, in this section we have introduced the $P$, $T$ and $C$ transformations.
Their commutation relations are
$$PT=TP,\quad CP=-CP,\quad TC=-CT.$$
We have also found the following relations
  \begin{gather*}
\begin{aligned}
  PH_\CB &=H_\CB P   &&\text{ if } A_l=A_{-l},\\
  TH_\CB &=H_\CB T   &&\text{ if } A_l, B_l, B'_l\in\NR,\\
  CH_\CB &=-H_\CB C   &&\text{ if } B_l=B_l',
\end{aligned}
  \end{gather*}
  which, if we restrict to the case of Hermitian Hamiltonian,
  take the more symmetric form
  \begin{gather*}
\begin{aligned}
  PH_\CB &=H_\CB P   &&\text{if } A_l\in\NR,\\
  TH_\CB &=H_\CB T   &&\text{if } A_l, B_l\in\NR,\\
  CH_\CB &=-H_\CB C   &&\text{if } B_l\in\NR.
\end{aligned}
\end{gather*}

  Finally, it is interesting to consider the combination of the three
  transformations, that we will denote by $\Theta$ for short,  that is $\Theta\equiv CPT$. Its matrix in the standard basis is
  $$(\Theta)=\begin{pmatrix}
  \ 0\ &\ \ii \mathbf{I}\ \\\ii \mathbf{I}&0
  \end{pmatrix}\overline{
    \phantom{\begin{pmatrix}
        \\\end{pmatrix}}
    }
    $$        
  which is a self-adjoint, antiunitary operator. Notice that except for a factor $\ii$ it acts on $\CB$ like the adjoint, i.e. $\Theta\xi=\ii\xi^\dagger$.
  If we compute the transformation of $H_\CB$ under $CPT$ we find
  $$(\Theta H_\CB \Theta)=
\begin{pmatrix}\mathbf{A}^+&\overline{\mathbf{B}'}\\
  -\overline{\mathbf{B}}&-(\mathbf{A}^+)^{\trans}
\end{pmatrix}=-(H_\CB)^+,
$$
and, therefore, if $H$ is Hermitian we always have
$${\Theta}H_\CB=-H_\CB {\Theta}$$
even if $H$ has none of the three discrete symmetries.
This is our form of the $CPT$ theorem \cite{CPT}.

\section{Zero modes}\label{sec:zero_modes}

In this section, we will show how the symmetries that we introduced
above determine the states of zero energy. We  will see that the
dimension of the space of zero modes characterizes  the universality class
of the corresponding Hamiltonian in a sense that will be made precise below.

We recall that the general Hermitian Hamiltonian that we will consider
in the sequel is
\begin{eqnarray*}
  H&=&\frac12\sum_{n=1}^N\sum_{l=-L}^L\hskip -1mm{\vphantom{\sum}}'
  \left(2 A_l\, a_n^\dagger a_{n+l}+
  B_l\, a_n^\dagger a_{n+l}^\dagger-\overline B_l\, a_na_{n+l}\right)\\
  &+&\con\,(a_1^\dagger a_N+a_N^\dagger a_1),
\end{eqnarray*}
with $A_{-l}=\overline A_l$ and $B_{-l}=-B_l$ and acts by the adjoint in the
space of Bogoliubov modes.

We decompose the Hamiltonian as the sum of the
unperturbed part (the Hamiltonian for $\con=0$) and the rest
$$H_\CB=H_0+\con V.$$
Besides, in order to achieve the thermodynamic limit, it will be convenient
to divide $H_0$ into the piece in the bulk and that in the edge.
By that we mean the following: first decompose the Hilbert space into
the subspace in the bulk, corresponding to those sites that for range
$L$ couplings do not interact with the defect,
and the rest, i.e. $\CB=\CB_\bulk \oplus \CB_\edge $ with

\hskip 1cm$\displaystyle \CB_\bulk =\lspan\{a^\dagger_{L+1},\dots,a^\dagger_{N-L},
a_{L+1},\dots,a_{N-L}\}$

\noindent
and

\hskip 1cm$\displaystyle \CB_\edge =\lspan\{a^\dagger_{1},\dots,a^\dagger_{L},
a^\dagger_{N-L+1},\dots,a^\dagger_{N},
a_1,\dots,a_{L},
a_{N-L+1},\dots,a_{N}
\}.
$

Now consider the orthogonal projectors associated to this decomposition 
$\Pi_\bulk $ and $\Pi_\edge $ and define $$H_\bulk =\Pi_\bulk  H_0
\quad {\rm and}\quad  H_\edge =\Pi_\edge  H_0.$$

In this section, we
are interested in the states of zero energy for $\con=0$. That is, we must
compute the kernel of $H_0$.
It is clear that $\Ker H_0=\Ker H_\bulk \cap \Ker H_\edge$.
We proceed to characterize $\CK=\Ker H_\bulk$. 

In the standard basis for $\CB$ and $\CB_\bulk$, the matrix of $H_\bulk$
can be written as
$$
(H_\bulk)=\begin{pmatrix}\mathbf{A}_\bulk&\mathbf{B}_\bulk\\
-\overline{\mathbf{B}}_\bulk&-\overline{\mathbf{A}}_\bulk\end{pmatrix},
$$
where
$\mathbf{A}_\bulk$, $\mathbf{B}_\bulk$ are $(N-2L)\times N$ dimensional
matrices given by
$$\mathbf{A}_\bulk=
\begin{pmatrix}
  A_{-L}&\cdots&A_0&\cdots&A_L&0&\cdots&\cdots&0\\
  0&A_{-L}&\cdots&A_0&\cdots&A_L&0&\cdots&0\\
  \vdots&\ddots&\ddots&&\ddots&&\ddots&\ddots&\vdots\\
  0&\cdots&0&A_{-L}&\cdots&A_0&\cdots&A_L&0\\
  0&\cdots&\cdots&0&A_{-L}&\cdots&A_{0}&\cdots&A_L
\end{pmatrix},
$$
and
$$                                                                              
\mathbf{B}_\bulk=                                                               
\begin{pmatrix}                                                                 
  B_{-L}&\cdots&0&\cdots&B_L&0&\cdots&\cdots&0\\                                
  0&B_{-L}&\cdots&0&\cdots&B_L&0&\cdots&0\\                                     
  \vdots&\ddots&\ddots&&\ddots&&\ddots&\ddots&\vdots\\                          
  0&\cdots&0&B_{-L}&\cdots&0&\cdots&B_L&0\\                                     
  0&\cdots&\cdots&0&B_{-L}&\cdots&{0}&\cdots&B_L                                
\end{pmatrix}.                                                                  
$$

Given the form of $(H_\bulk)$, one easily sees that, if the non
degeneracy condition $|A_L|^2\not=|B_L|^2$ is met, the rank of 
$(H_\bulk)$ is maximal, which implies $\dim\CK=4L$.
This is the first piece of information we need.

Now, let us consider the simplest case in which the chain enjoys $P$, $T$
and $C$ symmetries. This means that $H_\CB$ commutes with $P$ and $T$ and
anticommutes with $C$. It is equivalent to assuming that all the couplings
are real.

We must introduce one more transformation that we denote $\Gamma$
and acts on the standard basis in the following way
$$ \Gamma a^\dagger_n=\begin{cases}
\ a^\dagger_n& \mbox{ for } n\leq N/2\\
-a^\dagger_n& \mbox{ for } n > N/2
\end{cases},
\qquad
\Gamma a_n=\begin{cases}
\ a_n& \mbox{ for } n\leq N/2\\
-a_n& \mbox{ for } n > N/2
\end{cases}.
$$
That is, $\Gamma$ leaves invariant the first half of the chain
and reverts the sign of the second half.

It is clear that $\Gamma$ commutes with $H_\edge$ and
anticommutes with $V$. Besides, if the system has a gap between the valence and
conduction bands or, in other words, if the zero
energy states of $H_\bulk$ are localized at the edges with exponential decay,
then, in the thermodynamic limit, $\Gamma$ preserves $\CK$.
In the following section, when we define properly the thermodynamic limit,
we will make these statements more precise. 
For the moment, it is enough to assume that
$$\Gamma\CK\subset \CK.$$
From its definition, one has $\Gamma=\Gamma^\dagger=\Gamma^{-1}$ and
satisfies the following commutation relations
$$\Gamma P=-P\Gamma,\qquad\Gamma T=T\Gamma,\qquad\Gamma C=-C\Gamma.$$

Due to the symmetries of the system, $\CK$ is left invariant by $P$,
$C$ and $T$ transformations. From the commutation relations above, one may check
that $\Gamma$ and $\ii CP$ commute, hence we can classify the vector in the kernel of $H_\bulk$ according to the respective charges $+1$ or $-1$ with respect to each of both symmetries. Then, if we write the $\Gamma$ charge upstairs and that of $\ii CP$ downstairs, we have

$$\CK=\CK^+_+\oplus\CK^+_-\oplus\CK^-_+\oplus\CK^-_-,$$
and we denote by $n^+_+, n^+_-, n^-_+$
and $n^-_-$ the respective
dimensions. 

Note that $P$ or $C$ act inside $\CK$
and reverse both $\Gamma$ and the $\ii CP$ charges.
This implies that $n^+_+=n^-_-$ and $n^+_-=n^-_+$ and hence
$n^+_++n^+_- = n^-_++n^-_-= 2L$.

Now, in order to determine the zero modes of $H_0$,
we must consider the restriction to $\CK$ 
of the Hamiltonian at the {\it edge}
$$H_\edge:\CK\to\CB_\edge$$
and compute its kernel.

First, notice that $\CB_\edge$ is preserved by $P$, $C$, $T$ and  $\Gamma$
and therefore it can be decomposed into invariant subspaces under
the simultaneous action of the two commuting operators $\Gamma$ and $\ii CP$.
Using the same notation as before, we have
$$\CB_\edge=\CB^+_+\oplus\CB^+_-\oplus\CB^-_+\oplus\CB^-_-$$
and, given the definitions of the different operators, we find
$$\dim\CB^+_+=\dim\CB^+_-=\dim\CB^-_+=\dim\CB^-_-=L.$$

Second, we may use that $\Gamma$ commutes with $H_\edge$ while $\ii CP$
anticommutes, then we have
$$H_\edge:\CK^+_+\to\CB^+_-\,,\quad H_\edge:\CK^+_-\to\CB^+_+\,,\quad H_\edge:\CK^-_+\to\CB^-_-\,,\quad H_\edge:\CK^-_-\to\CB^-_+\,.$$

Therefore if, for instance, $n^+_+>L$ then the restriction of
$H_\edge$ to  $\CK^+_+$ has a kernel of dimension at least $n^+_+-L$
and the same is true when considering $\CK^-_-$ and $n^-_-$
that, as shown before, is equal to $n^+_+$. Given that $L=(n^+_++n^+_-)/2$
and taking into account the two subspaces, we finally get
$$\dim\Ker H_0\geq | n^+_+-n^+_-|,$$
where the absolute value has been introduced to cover the case
$n^+_->L$.
Our result here is a lower bound for the number of independent zero modes,
but using a genericity argument one sees that for {\it typical}
Hamiltonians the lower bound is saturated.
In the next section, we will explicitly show that this is indeed the case for
our family of Hamiltonians.

It is interesting to observe that 
the zero modes for a given Hamiltonian (at $\con=0$)
have all the same $\ii CP\Gamma$ charge:
$+1$ if $n^+_+>n^+_-$  or $-1$ in the opposite case.

In the next sections, we will compute explicitly the indices introduced above,
that determine the number of zero modes, we will also show that it is possible
to derive them as a topological charge and establish the bulk-edge
correspondence.

\section{Bogoliubov states}\label{sec:bogoliubov_states}

This section is devoted to the construction of the Bogoliubov modes
of the fermionic chain, i.e. the solutions of the eigenvalue equations
$$H_\CB\xi=-E\xi,\quad H_\CB\xi^\dagger=E\xi^\dagger.$$
So that in the particular case of $E=0$ we must recover the results of
the previous section.

We first consider the part of the Hamiltonian in the bulk,
and study the equation
$$H_\bulk\xi=-E\Pi_\bulk \xi.$$
It is immediate to see that we can solve it with the ansatz
\begin{equation}\label{ansatz}
  \xi_z=\sum_n (\alpha\; z^{n-1} a^\dagger_n +\beta\; z^{n-1} a_n)
  \end{equation}
where
\begin{equation}\label{eigenvalue}
\begin{pmatrix}
  E+\Phi(z)&\Xi(z)\\
  -\overline\Xi(z)&E-\overline\Phi(z)
\end{pmatrix}\begin{pmatrix}\alpha\\ \beta\end{pmatrix}=0.
\end{equation}
We have introduced the Laurent polynomials of degree $(-L,L)$ \cite{Ares2, Ares3}
  $$
  \Phi(z)=\sum_{l=-L}^L A_l z^l,\quad
  \Xi(z)=\sum_{l=-L}^L B_l z^l,
  $$
and   $\overline\Phi(z)$ in (\ref{eigenvalue}) stands for the Laurent polynomial
  in $z$ with complex conjugate coefficients. That is
  $$\overline\Phi(z)=\overline{\Phi(\overline z)}$$
  and analogously for $\overline\Xi(z)$.

  Using the properties of their coefficients (for Hermitian Hamiltonians),
  we can establish reflection relations when we replace the arguments $z$ by
  $z^{-1}$. Indeed, one has
  $$\Phi(z^{-1})=\overline\Phi(z),\quad\Xi(z^{-1})=-\Xi(z),$$
  which will be useful in the future.
  
  In order to have non trivial solutions for (\ref{eigenvalue}),
  the determinant of the matrix has to vanish,
  \begin{equation}\label{dispersion}
    \left( E+\Phi(z)\right)\left (E-\overline\Phi(z)\right)+
    \Xi(z)\overline\Xi(z)=0,
  \end{equation}
  and if we take $z=\ee^{\ii \theta}$ the expression above is, of course, the
  dispersion relation in implicit form.

  The latter can be easily solved for $E$ to get
  $$E_\theta=\pm\sqrt{\Phi_+^2(\ee^{\ii \theta})+|\Xi(\ee^{\ii \theta})|^2}-
  \Phi_-(\ee^{\ii \theta}),
    $$
  where
  $$\Phi_\pm(z)=\frac{\Phi(z)\pm\overline\Phi(z)}2$$
  are the even and odd part of $\Phi$ under inversion of its argument
  and both are real for $z=\ee^{\ii \theta}$. It is then clear that $E$
  is always real for $z=\ee^{\ii \theta}$.

  The possible values of $E_\theta$ determine the bands of positive and negative
  energy and, therefore, the gap between them is given by
  $$\Delta=2\min_\theta |E_\theta|.$$
The bands can be identified respectively with the conduction and valence band,
hence if $\Delta=0$ our system is a model for a conductor and if $\Delta>0$
it corresponds to an insulator.

  The latter is the situation we are interested in: when the gap opens
  there may appear Bogoliubov modes localized in the
  vicinity of the defect whose energy lays between the bands.
  As the value of the contact, $\con$, changes, the energy of the localized modes
  may vary from the valence band to the conduction one, allowing for the
  adiabatic pumping phenomenon. This is also analogous
  to what happens for topological insulators in higher dimensions,
  where states localized at the boundary interpolate between the bands
  rendering the material a conductor \cite{Asboth}.

  States with energy between the bands correspond to solutions of
  (\ref{dispersion}) with $|z|\not=1$. Notice that, if the non degenerate
  condition $|A_L|^2-|B_L|^2\not=0$ is met, the  Laurent polynomial in
  equation (\ref{dispersion}) has degree $(-2L,2L)$ and hence it
  is equivalent to a polynomial equation of degree $4L$,
  therefore it has exactly $4L$ complex solutions including multiplicities.
  Although there is no
  difficulty in dealing with the general case, with the aim of making 
  the exposition simpler, we will assume that all the roots are simple
  so we have $4L$ different solutions for (\ref{dispersion}).

  It will be useful to distinguish
  the roots with modulus greater than one, that we will denoted by $z^{\gra}_r$,
  from those which have modulus smaller than one, to be denoted by $z^{\peq}_r$.
  Due to the properties of the Laurent polynomials $\Phi$ and $\Xi$,
  it is clear that if $z$ is a solution of  (\ref{dispersion}) then
  $\overline z^{-1}$ is also a solution. This implies that we have the same
  number of roots of every type, exactly $2L$.

  The distinction of the two types of solutions is important because,
  for $z^{\peq}_r$ and due to its exponential decay, $\xi_{z^{\peq}_r}$ is supported, in the   thermodynamic limit, in the first half of the chain, from $1$ to $N/2$,
  while for the other roots we have $\xi_{z^{\gra}_r}$ supported in the second half.
  This fact can be expressed through the action of the operator $\Gamma$,
  so that we have
  $$\Gamma \xi_{z^{\peq}_r}=\xi_{z^{\peq}_r},\quad
  \Gamma \xi_{z^{\gra}_r}=-\xi_{z^{\peq}_r}.$$
  The general solution for the bulk equations is a linear combination of
  those for the different roots, hence
  the Bogoliubov mode has the form
$$\xi=\sum_{r=1}^{2L} (\lambda_r^{\peq} \xi_r^{\peq}+\lambda_r^{\gra} \xi_r^{\gra}),$$
where 
$$\xi_r^{\peq}=\sum_{n=1}^N (z_r^{\peq})^{n-1}(\alpha_r^{\peq} a_n^\dagger+
\beta_r^{\peq} a_n),\quad \xi_r^{\gra}=\sum_{n=1}^N (z_r^{\gra})^{n-N}(\alpha_r^{\gra} a_n^\dagger+
\beta_r^{\gra} a_n).$$
  
Now we have to consider the piece of the Hamiltonian near the  defect
$$(H_\edge+q V) \xi=-E\;\Pi_\edge\xi.$$
We can remove the dependence on $E$ by applying to $\xi$ the relations in
the bulk and using again the non degeneracy
condition we obtain, in the thermodynamic limit, the equivalent equations
for $\lambda_r^{\peq}$ and $\lambda_r^{\gra}$
\comment
    {
\begin{eqnarray}\label{local_eqs}
  \begin{split}
    &\sum_{r=1}^{2L}
    \lambda_r^{\peq} (\overline A_1 \alpha_r^{\peq}-B_1 \beta_r^{\peq})=
    \con\sum_{r=1}^{2L} \lambda_r^{\gra} \alpha_r^{\gra} (z_r^{\gra})^{-1},\quad
 \sum_{r=1}^{2L}
    \lambda_r^{\peq} (A_1 \beta_r^{\peq}-\overline B_1 \alpha_r^{\peq})=
    \con\sum_{r=1}^{2L} \lambda_r^{\gra} \beta_r^{\gra} (z_r^{\gra})^{-1}\\
    &\sum_{r=1}^{2L} \lambda_r^{\peq} \alpha_r^{\peq} (z_r^{\peq})^{-l}=0,\quad
    \sum_{r=1}^{2L} \lambda_r^{\peq} \beta_r^{\peq} (z_r^{\peq})^{-l}=0,\quad l=2,\dots,L\\
    &\sum_{r=1}^{2L} \lambda_r^{\gra} \alpha_r^{\gra} (z_r^{\gra})^{l}=0,\quad
\sum_{r=1}^{2L} \lambda_r^{\gra} \beta_r^{\gra} (z_r^{\gra})^{l}=0,\quad l=2,\dots, L\\
    &\sum_{r=1}^{2L}
    \lambda_r^{\gra} (A_1 \alpha_r^{\gra}+B_1 \beta_r^{\gra})=
    \con\sum_{r=1}^{2L} \lambda_r^{\peq} \alpha_r^{\peq} z_r^{\peq},\quad
 \sum_{r=1}^{2L}
     \lambda_r^{\gra} (\overline A_1 \beta_r^{\gra}+\overline B_1 \alpha_r^{\gra})=
    \con\sum_{r=1}^{2L} \lambda_r^{\peq} \beta_r^{\peq} z_r^{\peq}.
 \end{split}
\end{eqnarray}
}
\begin{eqnarray}\label{local_eqs}
  \begin{split}
    &\sum_{r=1}^{2L}
    \lambda_r^{\peq} \alpha_r^{\peq} (z_r^{\peq})^{-L}=
    \con\sum_{r=1}^{2L} \lambda_r^{\gra}
    \frac{A_L \alpha_r^{\gra}+B_L \beta_r^{\gra}}{|A_L|^2-|B_L|^2},\quad
    \sum_{r=1}^{2L}
    \lambda_r^{\peq} \beta_r^{\peq} (z_r^{\peq})^{-L}=
    \con\sum_{r=1}^{2L}  \lambda_r^{\gra}
    \frac{\overline A_L \beta_r^{\gra}+ \overline B_L \alpha_r^{\gra}}{|A_L|^2-|B_L|^2},
    \\
    &\sum_{r=1}^{2L} \lambda_r^{\peq} \alpha_r^{\peq} (z_r^{\peq})^{-l}=0,\quad
    \sum_{r=1}^{2L} \lambda_r^{\peq} \beta_r^{\peq} (z_r^{\peq})^{-l}=0,\quad l=1,\dots,L-1,
    \\
    &\sum_{r=1}^{2L} \lambda_r^{\gra} \alpha_r^{\gra} (z_r^{\gra})^{l}=0,\quad
    \sum_{r=1}^{2L} \lambda_r^{\gra} \beta_r^{\gra} (z_r^{\gra})^{l}=0,\quad l=1,\dots, L-1,
    \\
    &\sum_{r=1}^{2L}
    \lambda_r^{\gra} \alpha_r^{\gra} (z_r^{\gra})^{L}=
    \con\sum_{r=1}^{2L} \lambda_r^{\peq} 
    \frac{\overline A_L \alpha_r^{\peq}-B_L \beta_r^{\peq}}{|A_L|^2-|B_L|^2},\quad
    \sum_{r=1}^{2L}
    \lambda_r^{\gra} \beta_r^{\gra}(z_r^{\gra})^{L}=
    \con\sum_{r=1}^{2L}  \lambda_r^{\peq} 
    \frac{A_L \beta_r^{\peq}-\overline B_L \alpha_r^{\peq}}{|A_L|^2-|B_L|^2}.
 \end{split}
\end{eqnarray}

The previous, together with the equations in the bulk in (\ref{eigenvalue},\ref{dispersion})
for $z^\lessgtr_r$, $\alpha^\lessgtr_r$ and $\beta^\lessgtr_r$,
can be considered as the rigorous definition of the
spectrum of the localized modes in the thermodynamic limit
(notice that all the equations mentioned here are independent of $N$).

It is interesting to observe how $P$, $T$ and $C$ transformations
act in the space of localized states:
\begin{alignat*}{6}
  &P:&\,&z_r^{\peq}\mapsto (z_r^{\peq})^{-1}&\qquad&T:&\,&z_r^{\peq}\mapsto \overline{z_r^{\peq}}&\qquad
  &C:&\,&z_r^{\peq}\mapsto (z_r^{\peq})^{-1}\\
  &&&\alpha_r^{\peq}\mapsto \alpha_r^{\gra}&&&&\alpha_r^{\peq}\mapsto \overline{\alpha_r^{\peq}}&&&&\alpha_r^{\peq}\mapsto \ii\beta_r^{\gra}
\\
  &&&\beta_r^{\peq}\mapsto -\beta_r^{\gra}&&&&\beta_r^{\peq}\mapsto \overline{\beta_r^{\peq}}&&&&\beta_r^{\peq}\mapsto -\ii\alpha_r^{\gra}
\\
  &&&\lambda_r^{\peq}\mapsto \lambda_r^{\gra}&&&&\lambda_r^{\peq}\mapsto \overline{\lambda_r^{\peq}}&&&&\lambda_r^{\peq}\mapsto \lambda_r^{\gra}
\end{alignat*}
And analogously with the exchange of $>$ and $<$.
One easily checks that these transformations are symmetries of the equations
in (\ref{local_eqs})  for $A_l\in\NR$ ($P$),
for  $A_l, B_l\in\NR$ ($T$), and for $B_l\in\NR$ ($C$).

Finally, we have the action of the operator $\Gamma$,
\begin{alignat*}{6}
  &\Gamma:&\,&\lambda_r^{\peq}\mapsto -\lambda_r^{\peq}\\
  &&&\lambda_r^{\gra}\mapsto \lambda_r^{\gra}\\
  &&&\con\mapsto-\con,
\end{alignat*}
which always leaves the equations invariant.

Naturally the unknowns $\lambda^{\peq}_r$ and $\lambda^{\gra}_r$
are uncoupled when $\con=0$. In this case, the equations read simply
\begin{eqnarray}\label{local_eqs_0}
  \begin{split}
    &\sum_{r=1}^{2L} \lambda_r^{\peq} \alpha_r^{\peq} (z_r^{\peq})^{-l}=0,\quad
    \sum_{r=1}^{2L} \lambda_r^{\peq} \beta_r^{\peq} (z_r^{\peq})^{-l}=0,\quad l=1,\dots,L,
    \\
    &\sum_{r=1}^{2L} \lambda_r^{\gra} \alpha_r^{\gra} (z_r^{\gra})^{l}=0,\quad
    \ \sum_{r=1}^{2L} \lambda_r^{\gra} \beta_r^{\gra} (z_r^{\gra})^{l}=0,\quad\  l=1,\dots, L.
 \end{split}
\end{eqnarray}
Their solutions will be discussed in the next sections.

Now we are interested in the opposite situation $\con\to\pm\infty$.
One can show that the energy of the localized states in these limits
coincide with those for $q=0$. To do that one must consider the following ansatz 
$$\xi'=\sum_{r=1}^{2L} ({\lambda'}_r^{\peq} {\xi'}_r^{\peq}+{\lambda'}_r^{\gra} {\xi'}_r^{\gra}),$$
where
$${\xi'}_r^{\peq}=\sum_{n=2}^N (z_r^{\peq})^{n-2}(\alpha_r^{\peq} a_n^\dagger+
\beta_r^{\peq} a_n),\quad {\xi'}_r^{\gra}=\sum_{n=1}^{N-1} (z_r^{\gra})^{n-N+1}(\alpha_r^{\gra} a_n^\dagger+
\beta_r^{\gra} a_n).$$
The important point here is that the term for $n=1$
is missing in ${\xi'}_r^{\peq}$ and that for $n=N$ is absent in ${\xi'}_r^{\gra}$.

If we introduce this ansatz into the equation for the defect and use
$|A_L|^2-|B_L|^2\not=0$, we arrive when $\con\to\pm\infty$ at a set of equations
identical
to that for $\con=0$, namely
\begin{eqnarray}\label{local_eqs_infty}
  \begin{split}
    &\sum_{r=1}^{2L} {\lambda'}_r^{\peq} \alpha_r^{\peq} (z_r^{\peq})^{-l}=0,\quad
   \sum_{r=1}^{2L} {\lambda'}_r^{\peq} \beta_r^{\peq} (z_r^{\peq})^{-l}=0,\quad l=1,\dots,L,
    \\
    &\sum_{r=1}^{2L} {\lambda'}_r^{\gra} \alpha_r^{\gra} (z_r^{\gra})^{l}=0,\quad
    \ \sum_{r=1}^{2L} {\lambda'}_r^{\gra} \beta_r^{\gra} (z_r^{\gra})^{l}=0,\ \quad l=1,\dots, L.
 \end{split}
\end{eqnarray}
Therefore, the solutions are the same and the spectrum of localized states at
$q=0$ and $q=\pm\infty$ coincide. This fact is apparent when one examines the
numerical resolution  of the system.

\section{Bulk-edge correspondence}\label{sec:bulk_edge}

In this section, we shall apply the machinery presented above to rederive
in an analytic way the algebraic results obtained in Sec.~\ref{sec:zero_modes}.
With the new analytic tools we will be able to characterize the universality
classes of the system which depend on the symmetries preserved by the
deformations. 

First of all we will consider the zero modes of the
$C, P$ and $T$ symmetric chain at $\con=0$, as we did in Sec.~\ref{sec:zero_modes}.
In this case, the Laurent polynomials have real coefficients
and the solutions are of the form
$$\xi_{z}=\sum_n (\alpha\; z^{n-1} a^\dagger_n +\beta\; z^{n-1} a_n)$$
for the bulk Hamiltonian should satisfy
\begin{equation}\label{zero_energy}
\begin{pmatrix}
  \Phi(z)&\Xi(z)\\
  -\Xi(z)&-\Phi(z)
\end{pmatrix}\begin{pmatrix}\alpha\\ \beta\end{pmatrix}=0,
\end{equation}
while $z$ is a solution of
\begin{equation}\label{disper_fact}
(\Phi(z)+\Xi(z))(\Phi(z)-\Xi(z))=0.
\end{equation}

Now we may distinguish two kinds of roots depending on whether
it is a solution of the first factor, in which case it will be denoted by $z_+$, or of the second one denoted by $z_-$.
It is immediate to see that the solution for (\ref{zero_energy}) in
the first case satisfies $\alpha_+=\beta_+$ and the Bogoliubov mode is
{even} under the combined action of charge conjugation and parity
i.e. $\ii CP\xi_{z_+}={}\xi_{z_+}$, while for the second case
we have $\alpha_-=-\beta_-$ and $\ii CP\xi_{z_-}={-}\xi_{z_-}$.
On the other hand, recall from the previous section that
$\Gamma\xi_{z^{\peq}}=\xi_{z^{\peq}}$ for $|z^{\peq}|<1$ and
$\Gamma\xi_{z^{\gra}}=-\xi_{z^{\gra}}$ if $|z^{\gra}|>1$.

\begin{figure}
  \includegraphics[width=16cm]{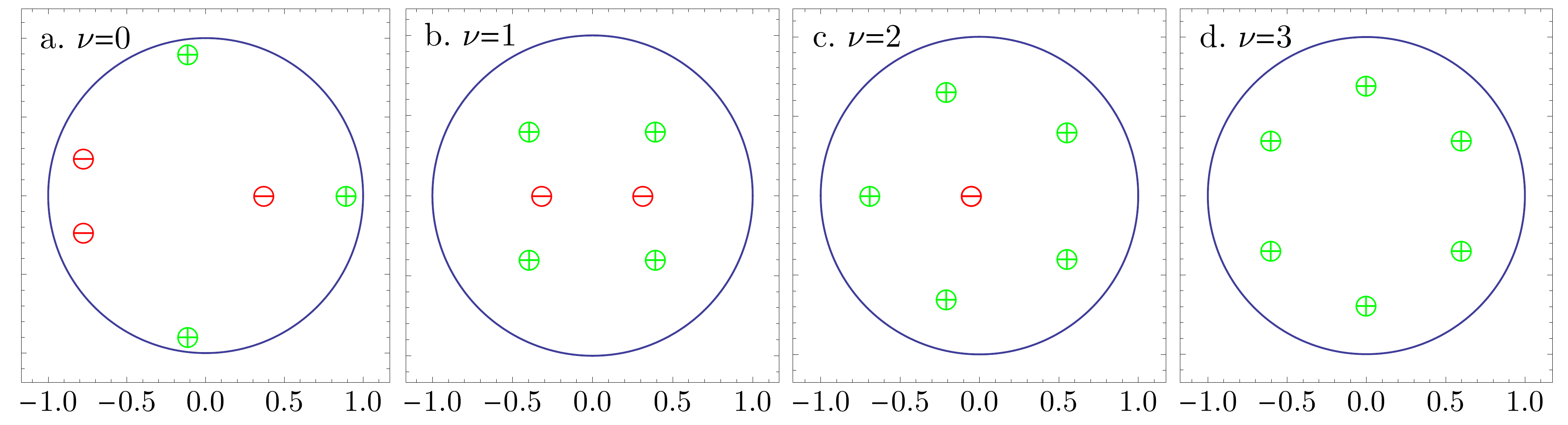}
\caption{
  In these plates we show the roots of (\ref{disper_fact}) inside the unit circle. The different plates
  correspond to Hamiltonians with different values of the topological 
  index $\nu$. Their non vanishing couplings are respectively: 
  (a) $A_0=0.9$, $A_1=1.0$, $A_3=-1.0$, $B_1=1.0$, $B_2=0.5$, $B_3=0.5$;
  (b) $A_1=0.5$, $B_1=0.5$, $B_3=-0.1$ ;
  (c) $A_2=0.5$, $A_3=0.1$, $B_2=0.5$, $B_3=-0.05$ and
  (d) $A_3=0.5$, $B_3=0.4$. 
   }
\label{fig:roots}
\end{figure}

Therefore, in order to determine the number
of zero modes for the open chain, it is enough to focus on
the roots of (\ref{disper_fact}) inside the unit circle and classify
them into those that make zero its first factor, $z^{\peq}_+$, and
those that annihilate the second one, $z^{\peq}_-$. Their respective numbers
are $n^+_+$ and $n^+_-$ using the notation of Sec.~\ref{sec:zero_modes}.

Now, according to the results in that section, the number of zero modes
is simply $|n^+_+-n^+_-|$. It is not difficult to rederive this result by
plugging into Eq.~(\ref{local_eqs_0}) our solutions
and examining the resulting system of equations. One immediately sees that the
subsystem for $\lambda^{\peq}_r$ has a kernel of dimension $\max(n^+_+,n^+_-)-(n_+^++n_-^+)/2$,
and the same for those for $\lambda^{\gra}_r$.
From this the result follows. In passing
we have shown that the bound for the limit on the dimension of
$\Ker H_0$ is saturated, as announced in Sec.~\ref{sec:zero_modes}.
In Fig.~\ref{fig:roots}, we represent the roots of \eqref{disper_fact} 
inside the unit circle for different values of the coupling constants 
and the index $\nu$. $\oplus$ stands for a root of the first 
factor of \eqref{disper_fact} and $\ominus$ for one of the second.

The number of zero modes as well as  the index $\nu=(n^+_+-n^+_-)/2$ are
topologically protected. The latter takes integer values from $-L$ to $L$,
and it can be written as a complex integral
(or a difference of the complex phase when traversing the Brillouin zone).
In concrete terms, consider the rational function
$$R(z)=\frac{\Phi(z)+\Xi(z)}{\Phi(z)-\Xi(z)}$$
then one has
$$\nu=\frac1{4\pi\ii}\oint_{|z|=1}\frac{R'(z)}{R(z)} \dd z$$
where the contour is travelled anticlockwise.

Now it is clear that, as long as the gap exists and the $C$, $P$ and $T$ symmetries are preserved, this number is topologically protected.
In fact, if we deform the Hamiltonian while keeping the symmetries,
the roots
$z^{\peq}_+$ and $z^{\peq}_-$ may move but they can not change suddenly their
character associated to the $\Gamma$ and $\ii CP$ charges.
The only way of varying the amount of solutions of each type
inside the disc is by letting the roots cross the unit circle,
but in that moment the gap closes (there are zero modes in the continuum
spectrum) and the topological protection associated to the existence of a gap
disappears.

Notice that our proof of the topological protection and the universality
classes requires that $P$ and $C$ are preserved so we have real
couplings. One may wonder what happens if these symmetries are absent.

In general, if the symmetries are lost, there are no zero modes at
$\con=0$ except for some exceptions which are still protected and will
be discussed later on.
But we may ask what is the fate of the universality classes.
The concrete question is the following: can we interpolate between two
$C$, $P$ and $T$ invariant theories with different topological
indices without closing the gap? The answer is affirmative provided the two
indices have the same parity, i.e. $\nu-\nu'=2k$. 
It is also true that in order to do so the interpolating Hamiltonians
must lose the $C$ symmetry.
The previous implies the existence of a $\NZ_2$ charge that is robust
under any perturbation that maintains the existence of a gap.  

We first show that if we preserve the $C$ symmetry in the intermediate
theories, then the topological index does not change. To see it, we first
notice that if $C$ is preserved, then the pairings $B_l$ are real
and the dispersion relation can be written as
$$ (E+\Phi_-(z))^2=\Phi_+^2(z)-\Xi^2(z).$$
Now if for some of the interpolating
Hamiltonians $\Phi_+^2(\ee^{\ii\theta})-\Xi^2(\ee^{\ii\theta})=0$ for some
$\theta$, then it is also so for $-\theta$ and, due to the properties of
$\Phi_-$ under inversion, $E_\theta=-E_{-\theta}$. This means that going
around the unit circle the sign of the energy changes, which implies
that it vanishes at some $\theta_0$ and the gap closes. If $\theta=-\theta$,
i.e.  if it is $0$ or $\pi$, one has that always
$\Phi_-(\pm 1)=0$ and $E_0=0$ or $E_\pi=0$ and the gap also closes.

Then the roots of $(\Phi_+(z)+\Xi(z))(\Phi_+(z)-\Xi(z))=0$
can not approach the boundary of the unit disc without closing the gap.
It implies that
$$ R(z)=\frac{\Phi_+(z)+\Xi(z)}{\Phi_+(z)-\Xi(z)}$$
is analytic around the unit circle for every intermediate Hamiltonian
and therefore the index defined from $R$ can not change at any step.

The second question is the invariance of the $\NZ_2$ topological charge under
any perturbation. In order to show that, we abandon for the moment the complex plane
and move to the Bogoliubov modes space.
In this context, we will prove that starting from a
$P$, $C$, and $T$ invariant Hamiltonian with odd $\nu$, 
then we necessarily have a pair of zero-energy states that does not disappear under
any perturbation of the couplings.

The reason is the following. As we saw in Sec.~\ref{sec:zero_modes},
in every $P$, $C$, $T$ invariant theory with gap
we can decompose the space of zero modes at $\con=0$
into eigenspaces of $\Gamma$,
$$\Ker H_0=\CV_0^+\oplus\CV_0^-$$
and $\dim\CV^\pm_0=|\nu|.$

Then the theory is deformed, call $H_t$ the deformed Hamiltonian,
in such a way that the gap does not disappear, i.e. we still
have states that are localized at the edge and its energy is preserved
under the transformation $\Gamma$.
The subspace $\Ker H_0$ is deformed
into a new $H_t$ invariant subspace $\CV_t$
that can be decomposed, as before, according to the eigenvalues of $\Gamma$, 
which still is a symmetry of the theory,
$$\CV_t=\CV_t^+\oplus\CV_t^-,$$
and both $\CV_t^+$ and $\CV_t^-$ are $H_t$ invariant.
Let us focus on $\CV_t^+$. Under a continuous deformation, its dimension
must be independent of $t$, therefore it is $|\nu|$.
Now let us consider a basis of $\CV_t^+$ formed by eigenstates
of $H_t$ and denote them by $\xi_E$. As
${\Theta}$ anticommutes with any Hermitian $H_t$, therefore
$${\Theta}\xi_E=\xi_{-E}$$
and, if all the states have $E\not=0$, then we must have an even number of them,
which implies that $|\nu|=\dim\CV^+_t$ is even. Or, reciprocally, if $\nu$ is odd
then there must be a state of zero energy invariant under
$CPT$. This shows the topological protection of the zero modes.
Of course, it ceases to be true when the gap closes and we lose
the symmetry under $\Gamma$.

It may be interesting to see this from a more constructive or geometrical
point of view that, besides, will give us an argument to prove that all
$P$, $C$ and $T$ symmetric theories with odd (even) $\nu$
are accessible by the continuous deformation of any of them,
maintaining the gap $\Delta>0$ in all the process. 

Consider a Hamiltonian with real couplings of range $L$.
The theory is fully determined by the $(-L,L)$ degree, Laurent polynomial
$$Q(z)=\Phi(z)+\Xi(z).$$
Actually we can single out $\Phi$ as the even part of $Q$
under inversion of $z$ and $\Xi$ as the odd part and
from them we can recover the
hopping and pairing constants $A_l$ and $B_l$ respectively.

$Q(z)$ is, in turn, determined up to a trivial scaling 
by its roots $z_r$. These, in order to have a gapped theory, should have
modulus different from one. By changing the $P$, $C$, $T$ invariant
Hamiltonian and as long as they do not cross the unit
circle (in which moment the theory would be gapless),  we can move the roots
freely  with the only restriction of being real numbers or
forming complex conjugate pairs. Then the only difficulty is how to interpolate
continuously between two theories whose polynomials $Q(z)$ have
a different number of roots
inside the unit disc. The latter is given by $L+\nu$, hence the task is
to interpolate between two Hamiltonians with different index $\nu$
or, in more physical terms, different number of zero modes.

Explicitly, consider the following $t$ dependent family of Laurent polynomials
$$Q_t(z)=\frac12(\ee^{-t} z+\ee^{t} z^{-1})P(z),\quad t\in[-1,1]$$
where $P(z)$ has degree $(-L+1,L-1)$ and real
coefficients. Assume that $P(z)$ is associated to a $L-1$ range Hamiltonian with a gap between the valence and conduction bands,
and denote its corresponding index by $\nu_0$ (that is, we assume that
$P(z)$ has $L+\nu_0-1$ roots inside the unit disc and $L-\nu_0-1$
outside).
Notice that if $z_1,\dots,z_{2L-2}$
are the roots of $P$, then those of $Q_t$ are
$z_1,\dots,z_{2L-2}, \ii\ee^t, -\ii\ee^t$. Therefore, the family of
polynomials interpolates between a theory with index $\nu_{-1}=\nu_0+1$
(or equivalently $L+\nu_0+1$ roots inside the unit disc for $t=-1$)
and another one with $\nu_1=\nu_0-1$ (or $L+\nu_0-1$ roots).

The problem with this interpolation is that the gap closes at $t=0$.
In fact, we have
$$\Delta_t^2=4\min_\theta|Q_t(\ee^{\ii\theta})|^2,$$
but for $t=0$ one has $Q_0(\ee^{\ii\pi/2})=0$, hence $\Delta_0=0$,
which is not allowed.

This inconvenience can be overcome if we consider a modification of the
interpolating Hamiltonians. If $Q_t$ is associated to
the couplings $A_l(t)$ and $B_l(t)$ we replace the pairing between
nearest neighbours $B_1(t)$ by
$$B'_1(t)=B_1(t)+\ii b \sqrt{1-t^2},$$
where $b$ is a constant to be appropriately
fixed. Notice that this new family of Hamiltonians
also interpolate between $Q_{-1}$ and $Q_1$, $B'(\pm1)=B(\pm1)$.
Next we will show that $H_t$ has a non zero gap for every $t\in[-1,1]$.

In fact, using the dispersion relation one can compute the new gap
$$\Delta'_t{}^2= 4\min_\theta\left((\cosh^2 t \cos^2\theta+\sinh^2 t \sin^2\theta)|P(\ee^{\ii\theta})|^2+4b^2(1-t^2)\sin^2\theta \right).$$
We have assumed that $P(\ee^{\ii\theta})\not=0$ for any $\theta$, then
there is a positive constant $a$ such that
$$|P(\ee^{\ii\theta})|^2> a^2 > 0,\ \mbox{for any}\ \theta\in[-\pi,\pi], $$
and if we choose $b=a/2$ we have
    \begin{eqnarray}
      \Delta'_t{}^2&\geq& 4a^2\min_\theta\left(\cosh^2 t \cos^2\theta+\sinh^2 t \sin^2\theta+(1-t^2) \sin^2\theta)\right)\cr
      &\geq&4a^2\min_\theta (1+ \sinh^2 t -t^2 +t^2\cos^2\theta)\cr
      &\geq & 4 a^2 (1+ \sinh^2 t -t^2) \geq 4 a^2.
    \end{eqnarray}
Therefore, throughout the whole process
the gap is bounded below by a positive constant which means that we
have constructed a legitimate, topological protective, interpolation.

It is interesting to pause a little and discuss why,
while this procedure can be used to connect a Hamiltonian
with index $\nu_0+1$ with another one with $\nu_0-1$,
we never could end up with a system of index $\nu_0$.

To understand why it is so, notice that if we move a
single root (instead of a pair of complex conjugate)
from inside the unit disc to its exterior
it should be real and cross the unit circle at $z=1$ or $z=-1$.
Now, due to its antisymmetric properties, $\Xi(\pm 1)=0$,
which implies that we can not open the gap at $z=\pm1$
with a modification of the couplings $B_l$,
as we did before (the same applies to $\Phi_-$, the antisymmetric
part of $\Phi$), and necessarily the interpolating
Hamiltonians have, at some point, zero gap. This is an
alternative explanation to the strong topological
protection of the $\NZ_2$ charge determined by the parity
of the index $\nu$.

\section{Charge pumping}\label{sec:pump}

In this section, we will study how the spectrum of the localized states
changes when we modify the value of the contact.
We will see that for certain Hamiltonians
there  are states that travel from the valence band to the
conduction one, providing a mechanism for the so called adiabatic
charge pumping.

\begin{figure}
  \includegraphics[width=16cm]{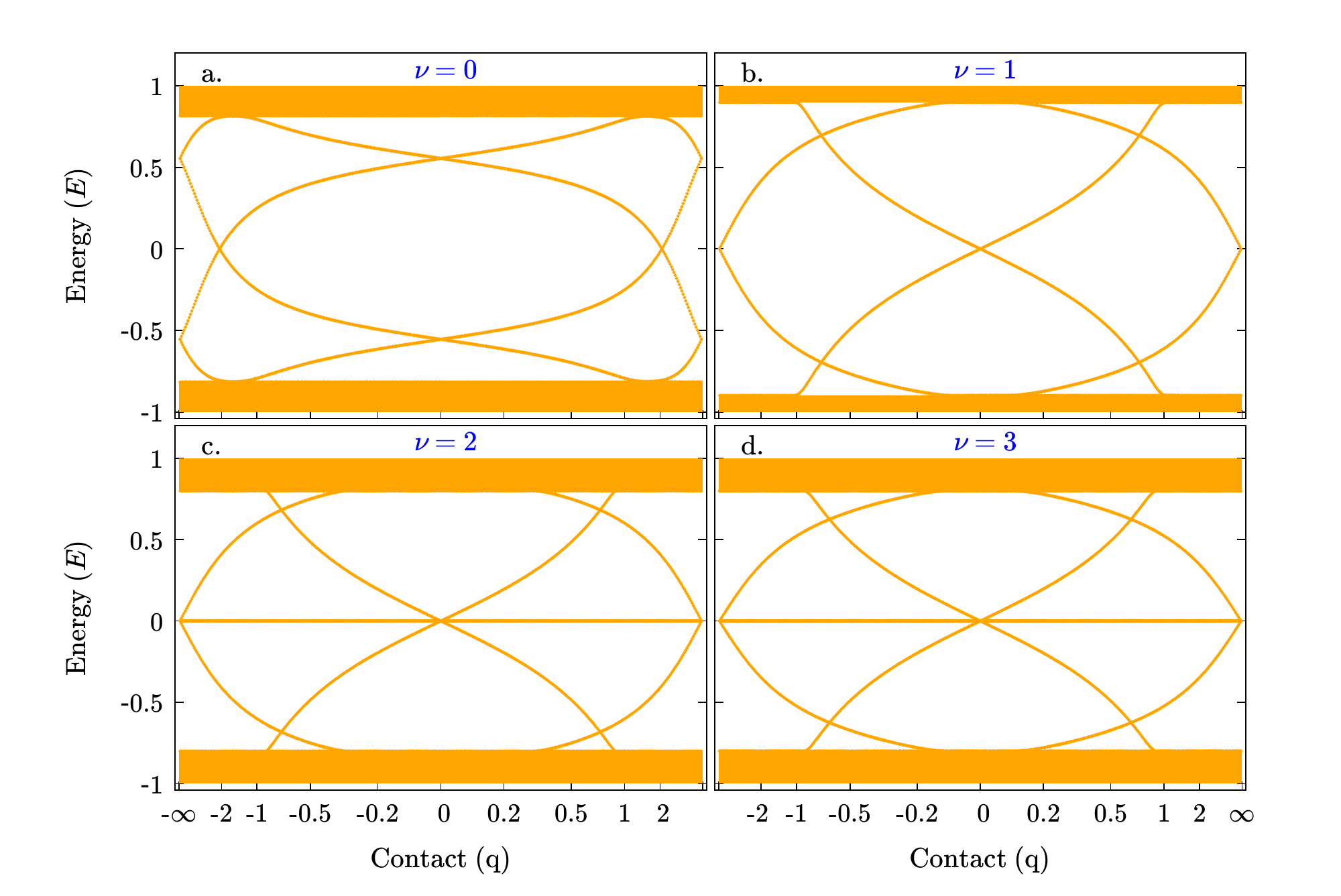}
\caption{
  In these plates we show part of the one particle spectrum that
  includes the gap, as a function of the contact. The different plates
  correspond to Hamiltonians with different values of the topological 
  index $\nu$. Their non vanishing couplings are respectively: 
  (a) $A_0=0.9$, $A_1=1.0$, $A_3=-1.0$, $B_1=1.0$, $B_2=0.5$, $B_3=0.5$;
  (b) $A_1=0.5$, $B_1=0.5$, $B_3=-0.1$ ;
  (c) $A_2=0.5$, $A_3=0.1$, $B_2=0.5$, $B_3=-0.05$ and
  (d) $A_3=0.5$, $B_3=0.4$. 
    The two figures in the lower row look similar, but they have the
  important difference that the horizontal line at $E=0$
  represents a doubly degenerate state in the plate (c) and
  a four times degenerate one in (d).}
\label{fig:four_classes}
\end{figure}

Let us illustrate this with an example. Consider the chain with topological index $\nu=0$ corresponding to plate a. in Fig. \ref{fig:four_classes}.
Assume that initially $q=-3$ and the system is at its minimum energy state,
i.e. we have all negative energy modes filled
(alternatively we can interpret this state as the vacuum of the Fock space $\ket{\Omega}$
given by $\xi_n\ket{\Omega}=0,\ n=1,\dots,N$). Now we proceed by slowly
increasing the value of $q$. At some point, approximately for $q=-2.053$,
one of the modes with negative energy crosses the $E=0$ level and produces
a hole in the Fermi sea. This represents the creation of a particle-hole
pair, but they are still localized at the defect. If we keep
adiabiatically increasing $q$ there is a moment (near $q=1.256$ in our case)
in which the mode of positive energy enters into the conduction band giving
rise to a free (unlocalized) particle-hole pair. This is what we mean by
the charge pumping phenomenon. 

We will see that all $P$, $C$, $T$ invariant theories enjoy the
charge pumping property, but their behaviour under perturbations
that break the symmetries differs from one universality class
to another.

For definiteness, we will consider the family of Hamiltonians with range
$L=3$, so that we have seven universality classes (if we restrict to the
$P$, $C$, $T$ invariant Hamiltonians), which correspond
to $\nu=0, \pm1, \pm2, \pm3$. Now, the theories with $\nu$ of opposite sign
and their perturbations are related by the change of $B_l$ to $-B_l$
that preserves the full spectrum. So in order to explore the whole zoo
of theories we may restrict to four classes with $\nu=0, 1, 2, 3$.

The spectrum of localized states, as a function of the contact $\con$, for a
representative of any of the four classes is plotted in Fig. \ref{fig:four_classes}.
At first sight, the plots corresponding to $\nu=2$ and $\nu=3$ look quite
similar. This is only apparent because, while in the case of $\nu=2$ the
states at zero energy that are present for any value of the contact
are doubly degenerate, those of $\nu=3$ are four times degenerate. This will
be manifest when we break the degeneracy by perturbing the respective
Hamiltonians with non symmetric terms.

{\bf Trivial case $\nu=0$.} 

In the first place, we shall consider the so called trivial
phase in which we have no edge states at zero energy in the open chain,
$\con=0$. As it was discussed in the previous sections, this occurs
in $P$, $C$ and $T$ symmetric theories with an index $\nu=0$.

\begin{figure}
  \includegraphics[width=16cm]{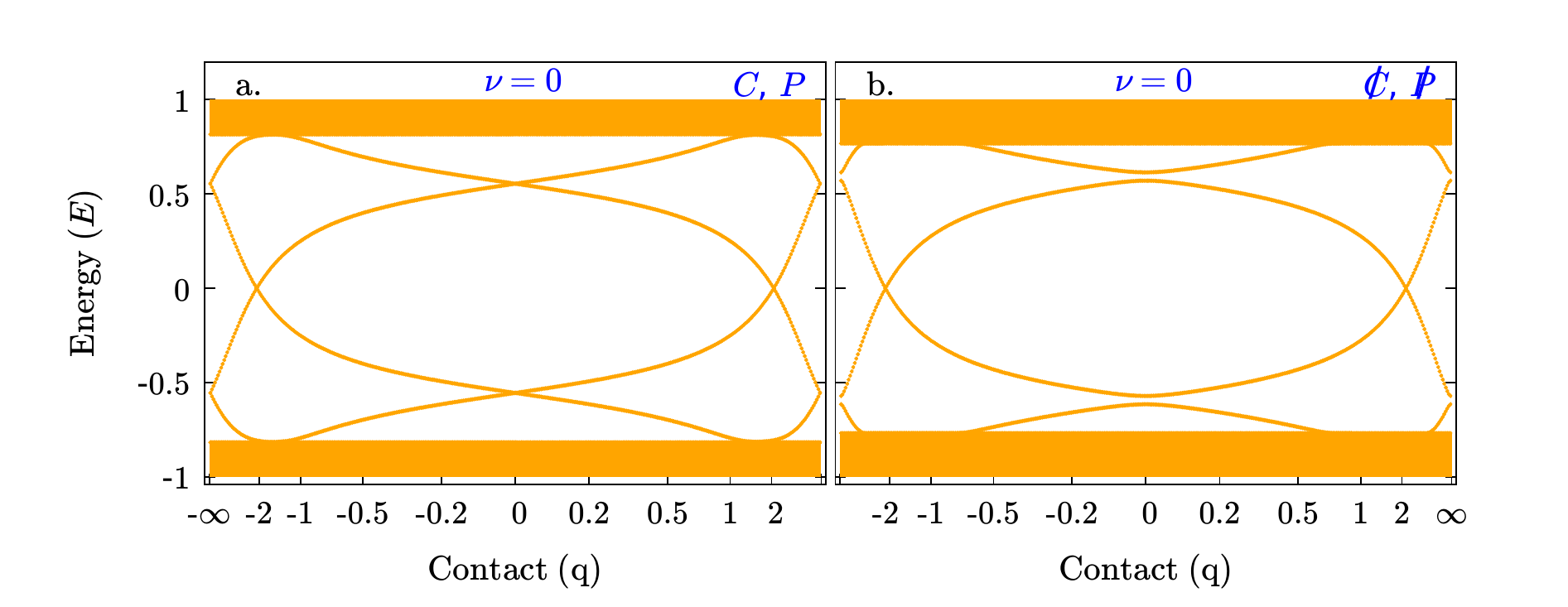}
  \caption{
    In the figure we show the spectrum  around the gap for two Hamiltonians
    in the trivial phase. Plate (a) is the same than the one in Fig.~\ref{fig:four_classes}
    and corresponds to a $C$ and $P$ symmetric Hamiltonian
    with couplings listed there. Plate (b)
    represents the spectrum for a deformation of the previous theory
    obtained by changing only $A_1=1.0+0.3\,\ii$ and $B_1=1.0+0.8\, \ii$.
    The addition of an imaginary part to the couplings breaks $C$ and $P$
    symmetries.
}
\label{fig:nu_0}
\end{figure}

The energy spectrum as a function of $\con$ is plotted in Fig. \ref{fig:nu_0}.
There we see that, while there are no zero modes at $\con=0$, 
we have them at some definite value of the contact.

We will show that this is a general property in the sense
that for our family of Hamiltonians there is always a zero energy
state at some value of $\con$.
In other words, we claim that there is always some value of
$\con$ such that
$$(H_0+\con V)\xi=0$$
has a solution with $\xi\not=0$.

To prove it, let us consider the subspace of $\CB$
containing the sites contiguous to the defect and 
denote it by $\CB_\vedge$. Explicitly
$$
\CB_\vedge =\lspan\{a^\dagger_{1},a^\dagger_{N},
a_1,a_{N}
\}
$$
and call $\Pi_\vedge$ the orthogonal
projector onto that space.

Of course, the relevant role played by $\CB_\vedge$
is due to the properties of the operator $V$, namely we
immediately see
$$\Pi_\vedge V= V\Pi_\vedge=V.$$
Note also that $\CB_\vedge$ is left invariant by the operators
$\Gamma$ and $\Theta$, that act on the standard basis respectively as
$$\Gamma a_1^\dagger=a_1^\dagger,\ \Gamma a_1=a_1,\ \Gamma a_N^\dagger=-a_N^\dagger,\ \Gamma a_N=-a_N,$$
$$\Theta a_1^\dagger=\ii a_1,\ \Theta a_1=\ii a_1^\dagger,\
\Theta a_N^\dagger=\ii a_N,\ \Theta a_N=\ii a_N^\dagger.$$

Now we have two possibilities: either $H_0$ has a non trivial kernel
(as it happens, for instance, with Hamiltonians in the universality class of odd $\nu$),
in which case
\begin{equation}\label{zeromode}
  (H_0+\con V)\xi=0
\end{equation}
has a non trivial solution for $\con=0$
or, if the previous does not hold, $H_0$ is invertible.

In this second case, we will consider the equation
\begin{equation}\label{zeromodepr}
  (V H_0^{-1}+\con^{-1}I)\xi'=0
\end{equation}
that is equivalent to (\ref{zeromode}) if $\con\not=0$
and we take $\xi'=H_0\xi$.

Of course, due to the properties of $V$, any solution
of (\ref{zeromodepr}) should belong to $\CB_\vedge$
and its equation can be equivalently written as
$$(V \Pi_\vedge H_0^{-1}|_{\CB_\vedge}+\con^{-1}I)\xi'=0.$$
Our next task is to determine the most general form of the map
$$\Pi_\vedge H_0^{-1}|_{\CB_\vedge}\,:\,\CB_\vedge\rightarrow\CB_\vedge.$$

In order to do that, we use the relations
$$H_0\Gamma=\Gamma H_0,\qquad H_0\Theta=-\Theta H_0$$
where, as argued earlier, the first one is true for
localized states and the second one holds for any Hermitian Hamiltonian.
These relations are also valid for its projected inverse.
Therefore, we can deduce the following properties for the matrix elements
of $\Pi_\vedge H_0^{-1}|_{\CB_\vedge}$
$$\prod{a_1^\dagger}{H_0^{-1}\, a_1^\dagger}=-\prod{a_1}{H_0^{-1}\, a_1}=\gamma_1\in\NR,$$
$$\prod{a_N^\dagger}{H_0^{-1}\, a_N^\dagger}=-\prod{a_N}{H_0^{-1}\, a_N}=\gamma_N\in\NR,$$
and all the others vanish.

From which we get the following expression for $V H_0^{-1}|_{\CB_\vedge}$
in the standard basis of $\CB_\vedge$
$$
(V H_0^{-1}|_{\CB_\vedge})=
\begin{pmatrix}
  0&1&0&0\cr
  1&0&0&0\cr
  0&0&0&-1\cr
  0&0&-1&0
\end{pmatrix}
\begin{pmatrix}
  \gamma_1&0&0&0\cr
  0&\gamma_N&0&0\cr
  0&0&-\gamma_1&0\cr
  0&0&0&-\gamma_N
\end{pmatrix}
=
\begin{pmatrix}
  0&\gamma_N&0&0\cr
  \gamma_1&0&0&0\cr
  0&0&0&\gamma_N\cr
  0&0&\gamma_1&0
\end{pmatrix}.
$$

Thus the eigenvalue
equation \pref{zeromodepr} has non trivial solutions if and only if
$$\det(V H_0^{-1}|_{\CB_\vedge}+\con^{-1} I)=(\con^{-2}-\gamma_1\gamma_N)^2=0.$$
We will show below that for our cases of interest (invertible $H_0$
continuously connected with a $P$ invariant Hamiltonian
without zero modes at $\con=0$)
$\gamma_1\gamma_N>0$ and hence there are two opposite real values of
$q=\pm(\gamma_1\gamma_N)^{-1/2}$ for which  \pref{zeromodepr} has non trivial
solutions or equivalently $H_0+qV$ has eigenstates localized at the defect
with zero energy.

First we will prove that $\gamma_1$ and $\gamma_N$ must be different from zero.
Let us assume the contrary and take for instance $\gamma_1=0$.
If we denote by $\Pi_\vedge^1$ the orthogonal projector onto
$$\CB_\vedge^1=\lspan\{a_1^\dagger,a_1\}$$
and take $\xi_0=H_0^{-1} a_1^\dagger$, then when $\gamma_1=0$ we have
$\xi_0\in(\CB_\vedge^1)^\perp$ and also
$$(I-\Pi_\vedge^1)H_0\xi_0=
(I-\Pi_\vedge^1)a_1^\dagger=0.$$
This means that the operator in $(\CB_\vedge^1)^\perp$
defined by $H_0'=(I-\Pi_\vedge^1)H_0|_{(\CB_\vedge^1)^\perp}$ has a zero
energy eigenstate given by $\xi_0$.
But the operator $H_0'$ is exactly like $H_0$ only in a chain in which
the site $1$ has been removed. In the thermodynamic limit
and for localized states, this suppression does not make any difference,
therefore it is contradictory that $H_0'$ has a localized state of
zero energy and $H_0$ has not. The consequence of this is that $\gamma_1,\gamma_N$
can not vanish.

To argue that $\gamma_1\gamma_N>0$ we use an argument of continuity. In fact,
for a $P$ invariant Hamiltonian $\gamma_1=\gamma_N$ and the positivity of the product
is guaranteed.
As we just showed, it can not vanish and therefore any deformation
of this system has also $\gamma_1\gamma_N>0$ and necessarily a state of zero
energy at some value of the contact $\con$, as we stated before.

We discussed and explained the similarities between the two plates in Fig.~\ref{fig:nu_0}, 
namely the existence of zero modes of the Hamitonian for some value of $\con$. As it is 
apparent, the main difference between them is the existence
of degenerate, localized states at $\con=0$ in the plate on the left, while
the degeneracy is broken in the plate on the right. Of course,
the difference is related to the breaking of $C$ and $P$ symmetries
as we will explain now.

Consider first a $P$ symmetric Hamitonian $H_{\CB}$,
then we have in particular $PH_0=H_0P$ and for localized states
also $\Gamma H_0=H_0\Gamma$. Then every $H_0$-eigenspace $\CS_E$
of localized states is left invariant by $P$
and can be obtained from eigenvectors of $\Gamma$.
But, as $\Gamma$ and $P$
anticommute, we have that if we take $\xi\in\CS_E$ with $\Gamma\xi=\xi$,
then $P\xi\in\CS_E$ and  $\Gamma P\xi=-P\xi$.
Therefore, $\xi$ and $P\xi$ are independent and $\CS_E$ has dimension
at least two, which explains the degeneracy of the energy level.

In the case that $C$ is not broken, we can repeat the previous argument
but replacing $P$ by $PT$ which, due to the
$CPT$ theorem, commute in this situation with $H_{\CB}$ and anticommutes
with $\Gamma$.

Finally, if both $C$ and $P$ are broken there is no reason for
the existence of degenerate states at $\con=0$
and, as it is shown in the plate on the right, the
two curves split.

Note in passing, that in this last case the charge pumping phenomenon
is not possible, as the adiabatic evolution fails to transfer
states from the valence band to the conduction one.

We finally would like to remark the coincidence of the
spectrum of localized states for $\con=0$ and $\con=\infty$.
This was explained in Sec.~\ref{sec:bogoliubov_states} and is clearly illustrated in the two
plates of Fig.~\ref{fig:nu_0}.

{\bf Nontrivial phase $\nu=1$.}

\begin{figure}
  \includegraphics[width=16cm]{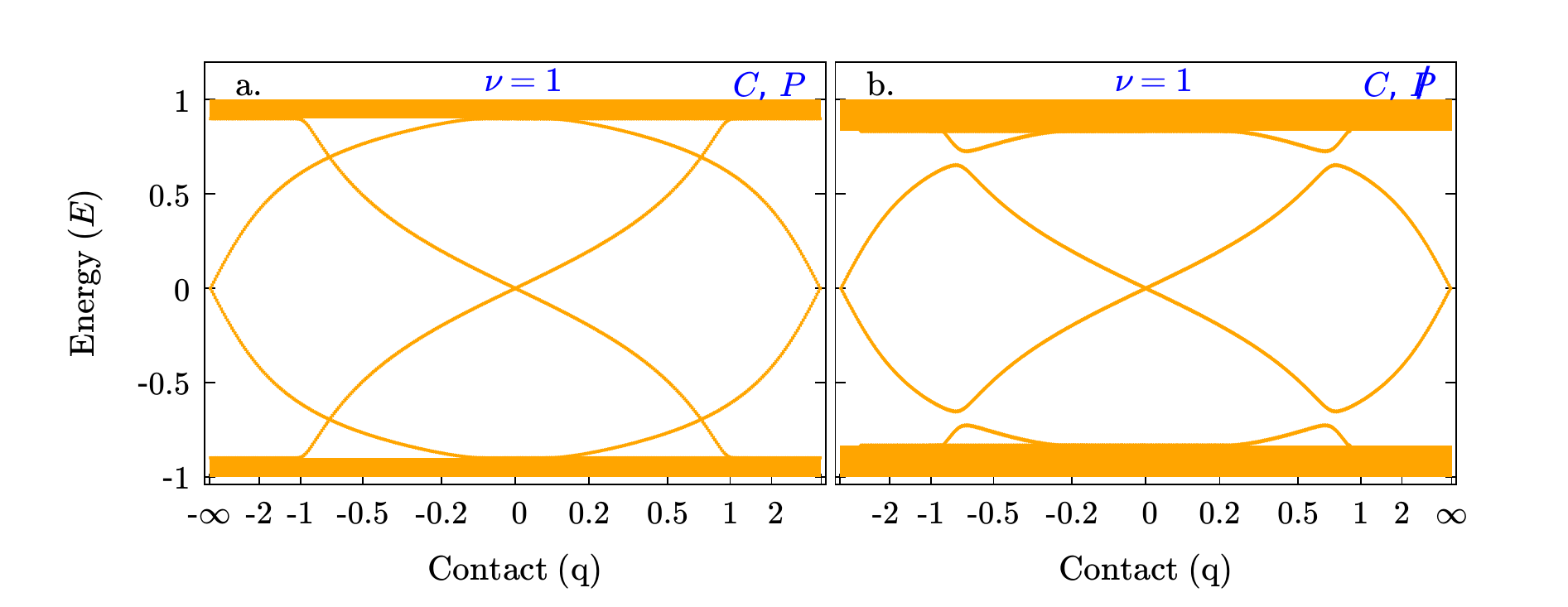}
  \caption{
    In the figure we show the spectrum  around the gap for a Hamiltonian
    with topological index $\nu=1$ and its deformation.
    Plate (a) is identical to that in Fig.~\ref{fig:four_classes} and corresponds to a
    $C$ and $P$ symmetric Hamiltonian with the coupling constants specified
    there.
    Plate (b) represents the spectrum for a deformation of the previous,
    obtained by changing the nearest neighbours hopping into
    $A_1=0.5+0.1\,\ii$, which breaks $P$ invariance.
}
\label{fig:nu_1}
\end{figure}

We move now to the case in which we have localized states of zero energy
for the open chain ($\con=0$). Two examples of this situation that correspond
to the topological index $\nu=1$ are plotted in Fig.~\ref{fig:nu_1}.

As we already discussed, and due to the fact that $\nu$ is odd, we can
not remove the zero modes at $\con=0$ by any continuous perturbation
of the system that maintains a gap between the bands along the whole process.
This is illustrated in Fig.~\ref{fig:nu_1}: in the left plate we depict the spectrum
of a $C$, $P$, $T$ symmetric chain while that of the right has $P$ and $T$
symmetries broken. We observe that in both cases there are two states
of zero energy for the open chain.

The main difference between the two plots is that in the left one we have
the possibility of pumping charge adiabatically from the valence to the
conduction band, while in the right one this is not possible.
In fact in the $C$, $P$, $T$ symmetric case, as we change $\con$, a state
moves continuously from one band to the other while in the non symmetric 
one there is a gap and there are not states that connect both
bands.

The two different regimes are determined by the properties
of the system under parity. One can show that if the chain is parity invariant
($PH_\CB=H_\CB P$), 
then there is necessarily a pair of states interpolating between the two bands.
To see this, let us denote by $\xi_\con$ a localized eigenstate for
$H_\CB=H_0+qV$ with energy $E_\con$
between the bands and assume
that it is even under parity, i.e. $P\xi_\con=\xi_\con$. Now due to
the Hellmann-Feynman theorem we have
$$\frac{\dd}{\dd\con}E_q=\prod{\xi_\con}{V \xi_\con}.$$
Then if we denote
$$\xi_\con=\sum_{n=1}^N(\alpha_n a^\dagger_n+\beta_n a_n)$$
and take into account the form of $V$ we have
$$\prod{\xi_\con}{V \xi_\con}=\overline\alpha_1\alpha_N+\alpha_1\overline\alpha_N
-\overline\beta_1\beta_N-\beta_1\overline\beta_N.$$
But $P\xi_\con=\xi_\con$ implies that $\alpha_1=\alpha_N$ and
$\beta_1=-\beta_N$. Therefore,
$$\prod{\xi_\con}{V \xi_\con}=2|\alpha_1|^2
+2|\beta_1|^2.
$$
Now if $\alpha_1\not=0$ or $\beta_1\not=0$ for any $\con$, then the energy as a
function of $\con$ has a positive slope and it interpolates between the
lower and the upper band. 

Before proceeding, let us pause a little to discuss the possibility
$\alpha_1=\beta_1=0$. In that case, we have $V\xi_q=0$ and
then $E_\con=E_0$ and $\xi_\con=\xi_0$, i.e. both the energy and the state
are independent of $\con$.
This would be represented in our plots by a horizontal line at the
given constant value of the energy $E_0$. It is also interesting to notice that
in the thermodynamic limit we can obtain two eigenstates $\xi'_0$
and $\Gamma\xi'_0$  with the same energy $E_0$ for $H_0$ by the
transformation $$\alpha'_1=\alpha_2,\ \ \alpha'_2=\alpha_3,\; \dots\;,\ \ 
\alpha'_{N-1}=\pm\alpha_{N-2},\  \ \alpha'_{N}=\pm\alpha_{N-1},
$$
where the $+$ and $-$ sign corresponds, respectively, to $\xi_0'$ and $\Gamma\xi_0'$.
To be more precise, and using the notation of Sec.~\ref{sec:bogoliubov_states}
where we give a rigorous definition of the thermodynamic limit,
$${\lambda^{\peq}_r}'=z^{\peq}_r{\lambda^{\peq}_r}
\quad\mbox{and}\quad {\lambda^{\gra}_r}'=\pm(z^{\gra}_r)^{-1}{\lambda^{\gra}_r}.$$
This is precisely the situation that we have for $E=0$ and $\con=0$ in the third
and fourth plates of Fig.~\ref{fig:four_classes} where, together with the horizontal line at
$E=0$, we have two oblique curves with parity $+$ and $-$ 
respectively, which correspond to the states $\xi'_0$ and $\Gamma\xi'_0$,
forming altogether a $\Cross$ shape. 
So far, this multiply degenerate situation for
stationary states at $q=0$ has been observed uniquely in these
particular cases ($P$, $C$ and $T$ symmetric theories with  $\nu\geq2$ and $E=0$). We do not know if they may appear elsewhere.

After this little digression, we proceed to conclude the analysis
of the differences between the two plates of Fig.~\ref{fig:nu_1}.
We already showed that for a $P$ invariant theory there are always
lines with a slope of constant sign in the plot of $E$ against $\con$.
This is not  necessarily so if parity is broken, as it is illustrated
in the plate on the right where the different curves in the region between
the bands reach a maximum or minimum and change the sign
of their slope. A consequence of this is that the connection between
the bands is interrupted and the phenomenon of charge pumping is not
possible any more. This is similar to what we observed in the trivial
($\nu=0$) case with the difference that there we had to break
both $C$ and $P$ symmetries to destroy the connection between the bands.

{\bf Nontrivial phases $\nu=2$ and $\nu=3$.}

We finally discuss the cases of higher topological index $\nu$
where the complexity of the $E-\con$ diagrams is shown in full lore.
Most of the features of the different plots and its relation
with the breaking of symmetries have been discussed already. Here
we will review how they actually manifest in these cases.

\begin{figure}
  \includegraphics[width=16cm]{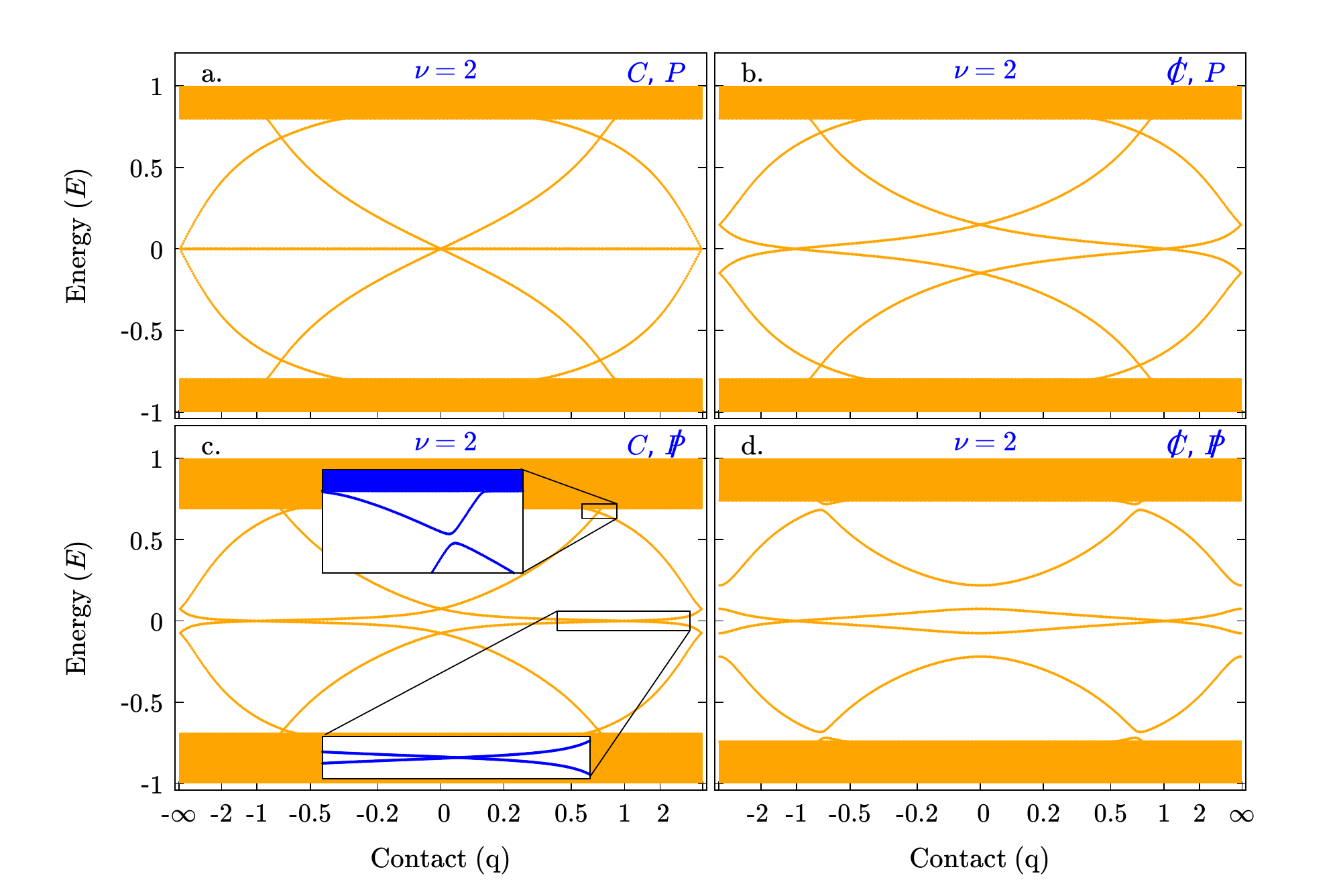}
  \caption{
    In these four plates we represent the spectrum around the gap for
    different deformations of the $C$, $P$, $T$ symmetric Hamiltonian
    corresponding to the third plot in Fig.~\ref{fig:four_classes} with topological index $\nu=2$.
    As listed there, the non vanishing couplings are
    $A_2=0.5$, $A_3=0.1$, $B_2=0.5$, $B_3=-0.05$ for the plate (a);
    we add to the previous Hamiltonian an imaginary nearest neighbour pairing
    $B_1=0.3\, \ii$, breaking $C$, in plate (b); a hopping
    $A_1=0.15\, \ii$, that breaks $P$ in plate (c) and, finally,
    we add both perturbations $A_1$ and $B_1$ together in plate (d).
  }
\label{fig:nu_2}
\end{figure}

If we start with the four plates of Fig.~\ref{fig:nu_2} corresponding to $\nu=2$,
we have in the upper-left the fully symmetric theory. As we already
mentioned, the horizontal line represents a doubly degenerate level.

The upper-right plot depicts the spectrum of a $P$ and non-$C$ symmetric theory.
We observe that the four degenerate states that we had  at
$\con=0$ and $E=0$ split into two pairs, as expected.
On the other hand, the adiabatic charge pumping is still
possible if we consider the compactification of the $\con$-axis
to form a circle.

In the lower-left plate, we show the spectrum for a $C$ and non-$P$
symmetric theory. Here the splitting of the four states in the middle
still occurs into pairs, but the connection between the bands and the
charge pumping phenomenon is destroyed due to the breaking of parity.

Finally, in the lower-right plate, all the symmetries are broken and
consequently there are not any degeneration,
except for the doubly degenerate states of zero energy
at opposite values of $\con$, whose necessary existence can be proven as we did in the $\nu=0$ case.
Notice also the coincidence of the spectrum at $\con=0$ and at $\con=\pm\infty$
in all four plates.

\begin{figure}
  \includegraphics[width=16cm]{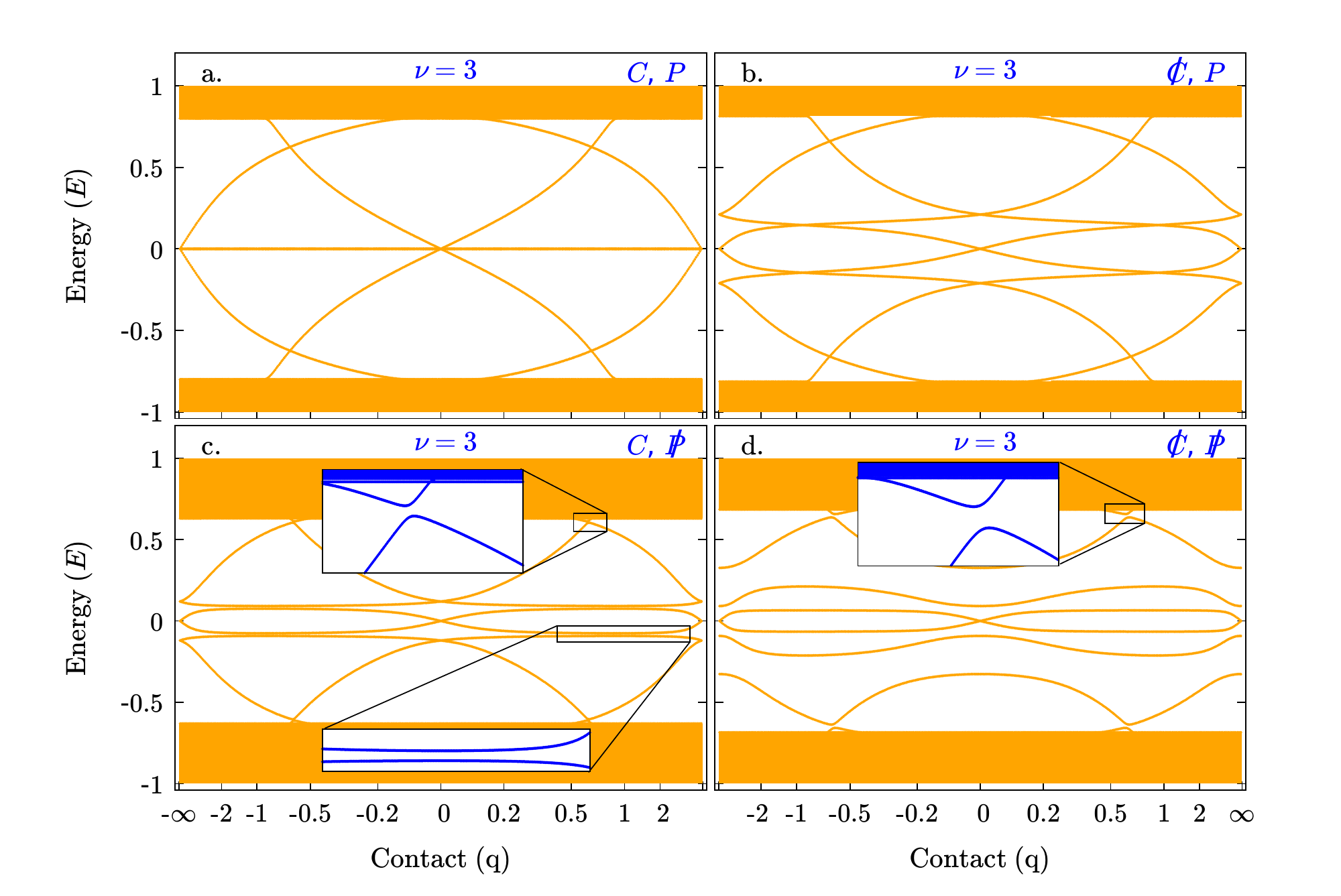}
  \caption{
    In these four plates we represent the spectrum around the gap for
    different deformations of the $C$, $P$, $T$ symmetric Hamiltonian
    corresponding to the fourth plot in Fig.~\ref{fig:four_classes} with topological index $\nu=3$.
    The spectrum of the undeformed Hamiltonian with the only
    non vanishing couplings given by 
    $A_3=0.5$, $B_3=0.4$ is represented in plate (a);
    the  previous, perturbed with an imaginary nearest neighbour pairing
    $B_1=0.3\, \ii$ breaking $C$ is plotted in plate (b); the original
    Hamiltonian supplemented with a hopping $A_1=0.17\, \ii$
    that breaks $P$ in plate (c) and, finally,
    the result of  adding both perturbations $A_1$ and $B_1$ together
    is shown in plate (d).
  }
\label{fig:nu_3}
\end{figure}

To conclude, let us briefly discuss the four plates of Fig.~\ref{fig:nu_3}.
They represent the spectrum of localized states for different
perturbations of a Hamiltonian with topological index  $\nu=3$.
As it is indicated in the plates, they differ by its behaviour
under $C$ and $P$ symmetries.
Here we see a situation very similar to the case $\nu=2$ (see Fig.~\ref{fig:nu_2}), but
a little more intricate.
The main difference with the previous case
is the persistence of a pair of zero modes at $q=0$ as it is expected according
to the arguments given in Sec.~\ref{sec:bulk_edge}.
Again, when the Hamiltonian is $P$ symmetric, we have the possibility
of adiabatic charge pumping from the valence band to the conduction one
and this disappears when the $P$ symmetry is broken.

\section{Conclusions}\label{sec:conclusions}

In this paper, we have studied the localized modes in a general free-fermionic 
chain with possible finite-range couplings and a single defect which breaks the 
translational invariance. The defect consists in a tunable hopping coupling $q$ that 
connects the end points of the chain. We have obtained a set of equations that allow 
to construct the edge modes for any values of the couplings and $q$ and can be directly extended
to the thermodynamic limit. 

For generic open chains ($q=0$) with $P$, $C$ and $T$ symmetries, we have determined the 
number of independent zero-energy modes, which characterizes the topological phase of the 
system. This analysis has been performed both algebraic and analytically. Through
the latter approach, we have derived a bulk-edge correspondence. The zero-modes can be associated with the roots of 
a polynomial. This polynomial is defined on the analytical continuation of the momentum space,
their coefficients are determined by the couplings of the Hamiltonian in the bulk and its 
degree by the the range $L$ of the couplings. Using this fact, we have introduced a
topological index which can be calculated by counting the roots of the polynomial and identifies 
the topological phases in agreement with the ten-fold classification \cite{Ryu}. 
We have found that the number of different topological phases depends on the range $L$, there are $2L+1$ phases
for $C$, $P$ and $T$ symmetric theories. It is interesting to point out that the 
analytical approach considered here is reminiscent of the geometric framework developed in 
Refs.~\cite{Its, Mezzadri, Ares2, Ares3} to analyze the entanglement entropies in this kind of systems.
This may have some connection with previous results that relate entropic and topological
properties of this kind of systems \cite{Preskill, Levin}.

When we turn on the contact $q$, we have found that the zero-energy modes may acquire energy, and
they can cross the gap of the bulk as $q$ varies, connecting the valence and conduction bands.
By pumping a localized state from one band to the other, one can create a free and delocalized particle-hole
pair. We have shown that this phenomenon occurs in any topological phase as long as the 
chain enjoys $P$ symmetry. In fact, the states that interpolate between bands have a defined parity.
In general, when we introduce perturbations in the couplings that break $P$ and $C$ symmetries, 
the connection between bands is lost. As we already mentioned in the introduction, an interesting aspect 
is that in our case we pump states by modifying adiabatically a parameter of the edge instead of one related to the 
bulk, as usually happens in the topological pumping.

Here we have only considered chains with finite long-range couplings, i.e.
the range $L$ of the couplings does not diverge when we take the thermodynamic
limit. But it would be interesting to extend the machinery and results 
presented in this paper to chains in which the couplings extend through the whole
chain, and the range $L$ diverges in the thermodynamic limit, such as the 
long-range Kitaev \cite{Vodola}. The presence of infinite-range 
couplings may give rise to unconventional features, see e.g. Refs.~\cite{Vodola2, Regemortel, Trombettoni, Ares4, Ares5}. In particular, novel 
topological phases and excitations can appear \cite{Viyuela, Alecce, Viyuela2, Lepori, Jager}. Another direction is to consider 
a non-Hermitian chain or defect, which can also host non-trivial and stable topological 
modes \cite{Martinez}. As in the case of infinite range couplings, the emergence of 
non-Hermitian topological phases implies the extension of the ten-fold way classification 
for Hermitian Hamiltonians \cite{Kawabata, Lieu} and the modification of the bulk-edge correspondence \cite{Kunst}. 
\newline

\textbf{Acknowledgments:} Research partially supported by grants E21\_17R, DGIID-DGA and 
PGC2018-095328-B-100, MINECO (Spain). FA acknowledges support from Brazilian Ministries 
MEC and MCTIC, from Simons Foundation (Grant Number 884966, AF), and from ERC under 
Consolidator Grant Number 771536 (NEMO), and acknowledges the warm hospitality and support 
of Departamento de F\'\i sica Te\'orica, Universidad de Zaragoza, during several stages of this work.


\begin{thebibliography}{XXX}

 \bibitem{Asorey} M. Asorey, \textit{Space, matter and topology},
 Nature Phys. 12, 616 (2016), arXiv:1607.00666 [cond-mat.mes-hall]

 \bibitem{Qi} X.-L. Qi, S.-C. Zhang, \textit{Topological insulators and superconductors}, 
 Rev. Mod. Phys. 83, 1057 (2011), arXiv:1008.2026 [cond-mat.mes-hall]
 
 \bibitem{Hasan} M. Z. Hasan, C. L. Kane, \textit{Colloquium: Topological insulators}, 
 Rev. Mod. Phys. 82, 3045 (2010), arXiv:1002.3895 [cond-mat.mes-hall]
 
 \bibitem{Bernevig}  A. Bernevig, T. L. Hughes, \textit{Topological Insulators and Topological Superconductors},
 Princeton: Princeton University Press, 2013
 
 \bibitem{KitaevMajorana} A. Kitaev, \textit{Unpaired Majorana fermions in quantum wires},
 Phys.-Usp. 44, 131 (2001), arXiv:cond-mat/0010440 [cond-mat.mes-hall]
 
 \bibitem{Rokhinson} L. P. Rokhinson, X. Liu, J. K. Furdyna, 
 \textit{Observation of the fractional ac Josephson effect: the signature of Majorana particles}, 
 Nature Phys. 8, 795 (2012), arXiv:1204.4212 [cond-mat.mes-hall]
 
 \bibitem{Das} A. Das, Y. Ronen, Y. Most, Y. Oreg, M. Heiblum, H. Shtrikman,
 \textit{Evidence of Majorana fermions in an Al-InAs nanowire topological superconductor},
 Nature Phys. 8, 887 (2012), arXiv:1205.7073 [cond-mat.mes-hall]
 
 \bibitem{Mourik} V. Mourik, K. Zuo, S. M. Frolov, S. R. Plissard, E. P. A. M. Bakkers, L. P. Kouwenhoven,
 \textit{Signatures of Majorana fermions in hybrid superconductor-semiconductor nanowire devices},
 Science 336, 1003 (2012), arXiv:1204.2792 [cond-mat.mes-hall]
 
 \bibitem{Nadj} S. Nadj-Perge, I. K. Drozdov, J. Li, H. Chen, S. Jeon, J. Seo, A. H. MacDonald, A. Bernevig, A. Yazdani,
 \textit{Observation of Majorana Fermions in Ferromagnetic Atomic Chains on a Superconductor},
 Science 346, 602 (2014), arXiv:1410.0682 [cond-mat.mes-hall]
 
 \bibitem{SarmaMaj} S. Das Sarma, H. Pan,
 \textit{Disorder-induced zero-bias peaks in Majorana nanowires},
 Phys. Rev. B 103, 195158 (2021), arXiv:2103.05628 [cond-mat.mes-hall]
 
 \bibitem{KitaevAnyons} A. Kitaev, \textit{Fault-tolerant quantum computation by anyons},
 Ann. Phys. 303, 2 (2003), arXiv:quant-ph/9707021
 
 \bibitem{Pachos} J. K. Pachos, \textit{Introduction to Topological Quantum Computation}, 
 Cambridge University Press, 2012
 
 \bibitem{Sarma} S. Das Sarma, M. Freedman, C. Nayak, \textit{Majorana Zero Modes and Topological Quantum Computation},
 npj Quantum Information 1, 15001 (2015), arXiv:1501.02813 [cond-mat.str-el]
 
 \bibitem{Altland} A. Altland, M. R. Zirnbauer,
 \textit{Nonstandard symmetry classes in mesoscopic normal-superconducting hybrid structures},
 Phys. Rev. B 55, 1142 (1997), arXiv:cond-mat/9602137
 
 \bibitem{Kitaev} A. Kitaev, \textit{Periodic table for topological insulators and superconductors}, 
 AIP Conf. Proc. 1134, 22 (2009), arXiv:0901.2686 [cond-mat.mes-hall]

 \bibitem{Ryu} S. Ryu, A. P. Schnyder, A. Furusaki, A. W. W. Ludwig, 
 \textit{Topological insulators and superconductors: tenfold way and dimensional hierarchy}, 
 New J. Phys. 12, 065010 (2010), arXiv:0912.2157 [cond-mat.mes-hall]
 
 \bibitem{RyuBB} S. Ryu, Y. Hatsugai,
 \textit{Topological Origin of Zero-Energy Edge States in Particle-Hole Symmetric Systems},
 Phys. Rev. Lett. 89, 077002 (2002), arXiv:cond-mat/0112197 [cond-mat.supr-con]
 
 \bibitem{Henkel} M. Henkel, A. Patk\'os, M. Schlottmann, \textit{The Ising quantum chain with defects. The exact solution},
 Nucl. Phys. B 314, 609 (1989)
 
 \bibitem{Grimm} U. Grimm, \textit{The quantum Ising chain with a generalized defect},
 Nucl. Phys. B 340 , 633 (1990), arXiv:hep-th/0310089
 
 \bibitem{Eisler} V. Eisler, I. Peschel, \textit{Solution of the fermionic entanglement problem with interface defects},
 Ann. Phys. 522, 679 (2010), arXiv:1005.2144 [cond-mat.stat-mech]
 
 \bibitem{Bertini} B. Bertini, M. Fagotti, \textit{Determination of the Nonequilibrium Steady State Emerging from a Defect},
 Phys. Rev. Lett. 117, 130402 (2016), arXiv:1604.04276 [cond-mat.stat-mech]
 
 \bibitem{Alase} A. Alase, E. Cobanera, G. Ortiz, L. Viola, \textit{Exact Solution of Quadratic Fermionic Hamiltonians for Arbitrary Boundary Conditions}, Phys. Rev. Lett. 117, 076804 (2016), arXiv:1601.05486 [cond-mat.supr-con]
 
 \bibitem{Cobanera} E. Cobanera, A. Alase, G. Ortiz, L. Viola, \textit{Generalization of Bloch's theorem for arbitrary boundary conditions:
 Interfaces and topological surface band structure}, Phys. Rev. B 98, 245423 (2018), arXiv:1808.07555 [cond-mat.stat-mech]
 
 \bibitem{Reyes-Lega} J. S. Calder\'on-Garc\'{\i}a, A. F. Reyes-Lega, \textit{Majorana Fermions and Orthogonal Complex Structures},
 Mod. Phys. Lett. A 33, 1840001 (2018), arXiv:1712.05069 [cond-mat.str-el]
 
 \bibitem{Najafi} M. N. Najafi, M. A. Rajabpour, \textit{Formation probabilities and statistics of observables as defect problems in free fermions and quantum spin chains}, Phys. Rev. B 101, 165415 (2020), arXiv:1911.04595 [cond-mat.stat-mech]
 
 \bibitem{Ares1} F. Ares, J. G. Esteve, F. Falceto, A. Us\'on, \textit{Complex behavior of the density in composite quantum systems}, 
 Phys. Rev. B 102, 165121 (2020), arXiv:2004.06813 [cond-mat.stat-mech]
 
 \bibitem{Thouless} D. J. Thouless, \textit{Quantization of particle transport},
 Phys. Rev. B 27, 6083 (1983)
 
 \bibitem{Asboth} J. K. Asb\'oth, L. Oroszl\'any, A. P\'alyi,
 \textit{A Short Course on Topological Insulators: Band-structure topology and edge states in one and two dimensions},
 Lecture Notes in Physics, 919 (2016), arXiv:1509.02295 [cond-mat.mes-hall]
 
 \bibitem{Kraus} Y. E. Kraus, Y. Lahini, Z. Ringel, M. Verbin, O. Zilberberg,
 \textit{Topological States and Adiabatic Pumping in Quasicrystals},
 Phys. Rev. Lett. 109, 106402 (2012), arXiv:1109.5983 [cond-mat.mes-hall]
 
 \bibitem{Lohse} M. Lohse, C. Schweizer, O. Zilberberg, M. Aidelsburger, I. Bloch, 
 \textit{A Thouless Quantum Pump with Ultracold Bosonic Atoms in an Optical Superlattice},
 Nature Phys. 12, 350 (2016), arXiv:1507.02225 [cond-mat.quant-gas]
 
 \bibitem{Nakajima} S. Nakajima, T. Tomita, S. Taie, T. Ichinose, H. Ozawa, L. Wang, M. Troyer, Y. Takahashi, 
 \textit{Topological Thouless Pumping of Ultracold Fermions}, 
 Nature Phys. 12, 296 (2016), arXiv:1507.02223 [cond-mat.quant-gas]
 
 \bibitem{Zilberberg} O. Zilberberg, S. Huang, J. Guglielmon, M. Wang, K. Chen, Y. E. Kraus, M. C. Rechtsman, \textit{Photonic topological pumping through the edges of a dynamical four-dimensional quantum Hall system}, 
 Nature 553, 59 (2018), arXiv:1705.08361 [quant-ph]
 
 \bibitem{Kuno} Y. Kuno, Y. Hatsugai, \textit{Interaction Induced Topological Charge Pump}, 
 Phys. Rev. Research 2, 042024 (2020), arXiv:2007.11215 [cond-mat.quant-gas]
 
 \bibitem{Teo} J. C. Y. Teo, C. L. Kane, \textit{Topological Defects and Gapless Modes in Insulators and Superconductors},
 Phys. Rev. B 82, 115120 (2010), arXiv:1006.0690 [cond-mat.mes-hall]
 
 \bibitem{Keselman} A. Keselman, L. Fu, A. Stern, E. Berg, \textit{Inducing time reversal invariant topological superconductivity and fermion parity pumping in quantum wires},
 Phys. Rev. Lett. 111, 116402 (2013), arXiv:1305.4948 [cond-mat.str-el]
 
 \bibitem{Lieb} E. Lieb, T. Schultz, D. Mattis, \textit{Two soluble models of an antiferromagnetic chain}, Ann. Phys. 16, 407 (1961)
 
 \bibitem{Uhlman} A. Uhlmann, \textit{Anti- (Conjugate) Linearity}, 
 Sci. China Phys. Mech. Astron. 59, 630301 (2016), arXiv:1507.06545 [quant-ph]
 
 \bibitem{CPT} R. F. Streater, A. S. Wightman, \textit{PCT, Spin and Statistics, and All That}, 
 Princeton Landmarks in Mathematics and Physics (2001)
 
 \bibitem{Its} A. R. Its, B.-Q. Jin, V. E. Korepin, \textit{Entanglement in XY Spin Chain}, 
 J. Phys. A: Math. Gen. 38, 2975 (2005), arXiv:quant-ph/0409027
 
 \bibitem{Mezzadri} A. R. Its, F. Mezzadri, M. Y. Mo, \textit{Entanglement entropy in quantum spin chains with finite range interaction},
 Commun. Math. Phys. 284, 117 (2008), arXiv:0708.0161 [math-ph]
 
 \bibitem{Ares2} F. Ares, J. G. Esteve, F. Falceto, A. R. de Queiroz, \textit{On the M\"obius transformation in the entanglement entropy of fermionic chains}, J. Stat. Mech. (2016) 043106, arXiv:1511.02382 [math-ph]
 
 \bibitem{Ares3} F. Ares, J. G. Esteve, F. Falceto, A. R. de Queiroz, \textit{Entanglement entropy and M\"obius transformations for critical fermionic chains}, J. Stat. Mech. (2017) 063104, arXiv:1612.07319 [quant-ph]
 
 \bibitem{Preskill} A. Kitaev, J. Preskill, \textit{Topological entanglement entropy}, 
 Phys. Rev. Lett. 96, 110404 (2005), arXiv:hep-th/0510092
 
 \bibitem{Levin} M. Levin, X.-G. Wen, \textit{Detecting topological order in a ground state wave function}, 
 Phys. Rev. Lett. 96, 110405 (2006), arXiv:cond-mat/0510613 [cond-mat.str-el]
 
 \bibitem{Vodola} D. Vodola, L. Lepori, E. Ercolessi, A. V. Gorshkov, G. Pupillo, \textit{Kitaev Chains with Long-Range Pairing},
 Phys. Rev. Lett. 113, 156402 (2014), arXiv:1405.5440 [cond-mat.str-el]
 
 \bibitem{Vodola2}  D. Vodola, L. Lepori, E. Ercolessi, G. Pupillo, \textit{Long-range Ising and Kitaev models: Phases, correlations and edge modes}, New J. Phys. 18,015001 (2016), arXiv:1508.00820 [cond-mat.str-el]
 
 \bibitem{Regemortel} M.  Van  Regemortel,  D.  Sels,  M.  Wouters,  \textit{Information propagation and equilibration in long-range Kitaev chains}, Phys. Rev. A 93, 032311 (2016), arXiv:1511.05459 [cond-mat.stat-mech]
 
 \bibitem{Trombettoni} L. Lepori, A. Trombettoni, D. Vodola, \textit{Singular dynamics and emergence of nonlocality in long-range quantum models}, J. Stat. Mech. 033102 (2017), arXiv:1607.05358 [cond-mat.str-el]
 
 \bibitem{Ares4} F. Ares, J. G. Esteve, F. Falceto, A. R. de Queiroz, \textit{Entanglement entropy in the long-range Kitaev chain}, 
 Phys. Rev. A 97, 062301 (2018), arXiv:1801.07043 [quant-ph]
 
 \bibitem{Ares5} F. Ares, J. G. Esteve, F. Falceto, Z. Zimboras, \textit{Sublogarithmic behaviour of the entanglement entropy in fermionic chains}, J. Stat. Mech. (2019) 093105, arXiv:1902.07540 [cond-mat.stat-mech]
 
 \bibitem{Viyuela} O. Viyuela, D. Vodola, G. Pupillo, M. A. Martin-Delgado, \textit{Topological Massive Dirac Edge Modes and Long-Range Superconducting Hamiltonians}, Phys. Rev. B 94, 125121 (2016), arXiv:1511.05018 [cond-mat.str-el]
 
 \bibitem{Alecce} A. Alecce, L. Dell'Anna, \textit{Extended Kitaev chain with longer-range hopping and pairing}, 
 Phys. Rev. B 95, 195160 (2017), arXiv:1703.10086 [cond-mat.str-el]
 
 \bibitem{Lepori} L. Lepori, L. Dell'Anna, \textit{Long-range topological insulators and weakened
bulk-boundary correspondence}, New J. Phys. 19, 103030 (2017), arXiv:1612.08155 [cond-mat.str-el]

 \bibitem{Viyuela2} O. Viyuela, L. Fu, M. A. Martin-Delgado, \textit{Chiral Topological Superconductors Enhanced by Long-Range Interactions},
Phys. Rev. Lett. 120, 017001 (2018), arXiv:1707.02326 [cond-mat.supr-con]

 \bibitem{Jager} S. B. J\"ager, L. Dell'Anna, G. Morigi, \textit{Edge states of the long-range Kitaev chain: an analytical study},
 Phys. Rev. B 102, 035152 (2020), arXiv:2006.00092 [cond-mat.str-el]
 
 \bibitem{Martinez} V. M. Martinez Alvarez, J. E. Barrios Vargas, M. Berdakin, L. E. F. Foa Torres, \textit{Topological states of non-Hermitian systems}, Eur. Phys. J. Spec. Top. 227, 1295 (2018), arXiv:1805.08200 [cond-mat.mes-hall]
 
 \bibitem{Lieu} S. Lieu, \textit{Topological phases in the non-Hermitian Su-Schrieffer-Heeger model}, 
 Phys. Rev. B 97, 045106 (2018), arXiv:1709.03788 [cond-mat.mes-hall]
 
 \bibitem{Kawabata} K. Kawabata, K. Shiozaki, M. Ueda, M. Sato, \textit{Symmetry and Topology in Non-Hermitian Physics},
 Phys. Rev. X 9, 041015 (2019), arXiv:1812.09133 [cond-mat.mes-hall]
 
 \bibitem{Kunst} F. K. Kunst, E. Edvardsson, J. C. Budich, E. J. Bergholtz, \textit{Biorthogonal Bulk-Boundary Correspondence in Non-Hermitian Systems}, Phys. Rev. Lett. 121, 026808 (2018), arXiv:1805.06492 [cond-mat.mes-hall]
 
\end{thebibliography}
\end{document}